\documentclass[aps]{revtex4}

\newcommand{\beq}{\begin{equation}}
\newcommand{\eeq}{\end{equation}}
\newcommand{\bear}{\begin{eqnarray}}
\newcommand{\ear}{\end{eqnarray}}
\newcommand{\earn}{\nonumber \end{eqnarray}}

\newcommand{\nn}{\nonumber \\}

\newcommand{\phisq}{\langle\phi^2\rangle}

\newcommand{\rc}{r_{\rm c}}
\newcommand{\mds}{m_{ \mbox{\tiny \sl DS}}}
\newcommand{\gsim}{\mathop{\lefteqn{\raise.9pt\hbox{$>$}}
\raise-3.7pt\hbox{$\sim$}}}
\newcommand{\lsim}{\mathop{\lefteqn{\raise.9pt\hbox{$<$}}
\raise-3.7pt\hbox{$\sim$}}}
\arraycolsep=\mathsurround

\begin{document}

\title{Analytical approximation of the stress-energy tensor
of a quantized scalar field in static spherically symmetric
spacetimes}

\author{Arkady A. Popov\thanks{Email address: popov@kspu.kcn.ru}}

\address{Department of Mathematics, Kazan State Pedagogical University,
Mezhlauk 1 st., Kazan 420021, Russia}

\begin{abstract}
Analytical approximations for ${\langle \varphi^2 \rangle}$ and
${\langle T^{\mu}_{\nu} \rangle}$ of a quantized scalar field in
static spherically symmetric spacetimes are obtained. The field is
assumed to be both massive and massless, with an arbitrary
coupling $\xi$ to the scalar curvature, and in a zero temperature
vacuum state.  The expressions for ${\langle \varphi^2 \rangle}$
and ${\langle T^{\mu}_{\nu} \rangle}$ are divided into low- and
high-frequency parts. The contributions of the high-frequency
modes to these quantities are calculated for an arbitrary quantum
state. As an example, the low-frequency contributions to ${\langle
\varphi^2 \rangle}$ and ${\langle T^{\mu}_{\nu} \rangle}$ are
calculated in asymptotically flat spacetimes in a quantum state
corresponding to the Minkowski vacuum (Boulware quantum state).
The limits of the applicability of these approximations are
discussed.

\vskip12pt\noindent {\normalsize PACS number(s): 04.62.+v,
04.70.Dy}
\end{abstract}

\maketitle

\section{Introduction}

The interest in the vacuum polarization effects in strong
gravitational fields is connected mainly with investigations of
the early Universe and the construction of a self-consistent model
of black hole evaporation. The main objects to calculate from
quantum field theory in curved spacetime are the quantities
$\left< \varphi^2 \right>$ and $\left< T^{\mu}_{\nu} \right>$
where $\varphi$ is the quantum field and $T^{\mu}_{\nu}$ is the
stress-energy tensor operator for $\varphi$. The latter quantity
is of particular interest as a source term in the semiclassical
Einstein field equations:
     \beq
     G^{\mu}_{\nu}=8 \pi G \left< T^{\mu}_{\nu} \right>.
     \eeq
Except for very special spacetimes, on which quantum matter fields
are propagated, and for boundary conditions with a high degree of
symmetry (see, for example, \cite{DC}), it is not possible to
obtain exact expressions for these quantities. Numerical
computations of these quantities are as a rule extremely intensive
\cite{HC,C,AHS,BBK}. Thus it is useful, when possible, to have
analytical approximations to ${\langle \varphi^2 \rangle}$ and
${\langle T^{\mu}_{\nu} \rangle}$. One of the most widely used
techniques to obtain information about these quantities is the
DeWitt-Schwinger expansion \cite{DW}. It may be used to give the
expansions for ${\langle \varphi^2 \rangle}$ and ${\langle
T^{\mu}_{\nu} \rangle}$ in terms of powers of $m L$ where $m$ is
the mass of the quantized field and $L$ is the characteristic
scale of change of the background gravitational field. For
conformally coupled massless fields approximate calculations have
also been made. For ${ \langle T^{\mu}_{\nu}\rangle }$ in static
Einstein spacetimes ($R_{\mu \nu}=\Lambda g_{\mu \nu}$) these
include the approximations of Page, Brown, and Ottewill
\cite{Page,BO,BOP}. These results have been generalized to
arbitrary static spacetimes by Zannias \cite{Z}. A different
approach to the derivation of approximate expressions for
${\langle \varphi^2 \rangle}$ and ${\langle T^{\mu}_{\nu}
\rangle}$ for conformally coupled massless fields in static
spacetimes has been proposed by Frolov and Zel'nikov \cite{FZ}.
Their calculations were based primarily on geometric arguments and
the common properties of the stress-energy tensor rather than on a
field theory. Using the methods of quantum field theory the
expressions for ${\langle \varphi^2 \rangle}$ and ${\langle
T^{\mu}_{\nu} \rangle}$ of a scalar field in static spherically
symmetric asymptotically flat spacetimes have been obtained by
Anderson, Hiscock, and Samuel \cite{AHS}. They assumed that the
field is massive or massless with an arbitrary coupling $\xi$ to
the scalar curvature and in a zero temperature quantum state or a
nonzero temperature thermal state. The result was presented as a
sum of two parts, numerical and analytical:
      \beq
      \langle T^{\mu}_{\nu} \rangle_{ren}=\langle T^{\mu}_{\nu}
      \rangle_{numeric}+\langle T^{\mu}_{\nu} \rangle_{analytic}.
      \eeq
The analytical part of their expression is conserved. This has a
trace equal to the trace anomaly for the conformally invariant
field. For these reasons they proposed to use $\langle
T^{\mu}_{\nu} \rangle_{analytic}$ directly as an approximation for
$\langle T^{\mu}_{\nu} \rangle_{ren}$. An analogous result has
been obtained by Groves, Anderson, and Carlson \cite{GAC}  in the
case of a massless spin $\frac{1}{2}$ field in a general static
spherically symmetric spacetime.

There are two questions about all of these approximations. First,
what are the limits of applicability of these approximations? And
second, how can one describe quantum states not considered in the
above mentioned works?

The DeWitt-Schwinger expansion is independent of the quantum
state. The criterion for the validity of the DeWitt-Schwinger
approximation is well known: $m L \gg 1$. The validity of the
other approximations discussed above was investigated by the
authors by means of comparison with earlier known results.

In this paper, approximate expressions for ${\langle \varphi^2
\rangle}_{ren}$ and ${\langle T^{\mu}_{\nu} \rangle}_{ren}$ of a
quantized scalar field in static spherically symmetric spacetimes
are derived. The field is assumed to be both massless or massive
with an arbitrary coupling $\xi$ to the scalar curvature $R$, and
in a zero temperature vacuum state. The expressions for ${\langle
\varphi^2 \rangle}_{ren}$ and ${\langle T^{\mu}_{\nu}
\rangle}_{ren}$ are divided into low- and high-frequency parts.
The Anderson-Hiscock-Samuel approach \cite{AHS} is used for
derivation of the high-frequency contributions to these
quantities. Being an ultraviolet effect, the high-frequency
contribution is general in the sense that its value does not
depend on the quantum state in which ${\langle \varphi^2
\rangle}_{ren}$ and ${\langle T^{\mu}_{\nu} \rangle}_{ren}$ are
taken. This contribution contains all the ultraviolet divergences
and can be renormalized. The low-frequency contributions are
determined by the quantum state and, as an example, such
contributions are calculated in asymptotically flat spacetime in a
quantum state corresponding to the Minkowski vacuum  (in generally
accepted terminology this corresponds to the choice of the
Boulware quantum state). Both parts of ${\langle T^{\mu}_{\nu}
\rangle}_{ren}$ are separately conserved. For a conformally
invariant field the trace of the high-frequency part of ${\langle
T^{\mu}_{\nu} \rangle}_{ren}$ is equal to the trace anomaly and
that of the low-frequency part of ${\langle T^{\mu}_{\nu}
\rangle}_{ren}$ is equal to zero. The validity of the
approximations is discussed.

In Sec. II the expressions for the Euclidian Green's function and
the unrenormalized stress-energy tensor of a scalar field with
arbitrary mass and curvature coupling in a general static
spherically symmetric spacetime are derived. In Sec. III the WKB
approximations for ${\langle \varphi^2 \rangle}$ and ${\langle
T^{\mu}_{\nu} \rangle}$ of a very massive field are discussed.
Section IV describes the method of calculating the high-frequency
contributions to these quantities and the renormalization
procedure for ${\langle \varphi^2 \rangle}$ and ${\langle
T^{\mu}_{\nu} \rangle}$. The low-frequency contributions in the
case of asymptotically flat spacetimes are derived and the
renormalized expressions for ${\langle \varphi^2 \rangle}$ and
${\langle T^{\mu}_{\nu} \rangle}$ are displayed in Sec. V. The
results are summarized in Sec. VI. In the Appendixes some
technical results are derived and long expressions are displayed.

The units $\hbar=c=1$ are used throughout the paper.


\section{An unrenormalized expression for $\left<
\varphi^2 \right>$ and $\left< T^{\mu}_{\nu} \right>$}

In this section the main points of the Anderson-Hiscock-Samuel
approach \cite{AHS} for obtaining unrenormalized expressions for
$\left< \varphi^2 \right>$ and $\left< T^{\mu}_{\nu} \right>$ are
outlined.

First of all the metric of the static spherically symmetric
spacetime is analytically continued into Euclidean form
  \beq\label{metric}
  ds^2= fd\tau^2+d\rho^2+r^2(d\theta^2+\sin^2\theta\, d\varphi^2),
  \eeq
where $f$ and $r$ are functions of $\rho$, and $\tau$ is the
Euclidean time ($\tau = it,$ where $t$ is the coordinate
corresponding to the timelike Killing vector, which always exists
in static spacetime).

The regularization of $\left< \varphi^2 \right>$ and $\left<
T^{\mu}_{\nu} \right>$ is achieved by using the method of point
splitting. When the points are separated one can show that
        \bear \label{phi2}
        \langle \varphi^2 \rangle_{unren}&=&
        G_E(x,\tilde x),
        \ear
        \bear \label{Tmnunren}
        \langle T^{\mu}_{\nu} \rangle_{unren}&=&
        \left( {1/2-\xi} \right) (g^{\mu \tilde \alpha}
        G_{E; \tilde \alpha \nu} +
        g_{\nu}^{\tilde \alpha} G_{E;}{}^{\mu}{}_{\tilde \alpha}) +
        (2 \xi-1/2) \delta^{\mu}_{\nu} g^{\sigma \tilde \alpha}
        G_{E; \sigma \tilde \alpha}
        -\xi (G_{E;}{}^{\mu}{}_{\nu} \nonumber \\ &&
        +g^{\mu \tilde \alpha}g^{\tilde \beta}_{\nu}
        G_{E;\tilde \alpha \tilde \beta})
        +2 \xi \delta^{\mu}_{\nu} (m^2 + \xi R) G_E
        +\xi (R^{\mu}_{\nu}-\delta^{\mu}_{\nu} R / 2)
        G_E-\delta^{\mu}_{\nu} m^2 G_E/2,
        \ear
where $m$ is the mass of the scalar field, $\xi$ is its coupling
to the scalar curvature $R$, and $g^{\tilde \alpha}_{\beta}$ is
the bivector of parallel transport of a vector at $\tilde x$ to
one at $x$.

The integral representation for the Euclidean Green's function
$G_E(x,\tilde x)$ of a scalar field in a static spherically
symmetric spacetime used by Anderson \emph {et al.} \cite{AHS} is
the following:
      \bear \label{GE}
      G_E(x;\tilde x)=\frac{1}{4 \pi^2}\int \nolimits_{0}^{\infty } d \omega
      \cos [\omega (\tau - \tilde \tau)] \sum \limits_{l=0}^{\infty}
      (2l+1) P_l (\cos \gamma ) \ C_{\omega l} \ p_{\omega l}(\rho_<)
      \ q_{\omega l}(\rho_>),
      \ear
where $P_l$ is a Legendre polynomial, $\cos \gamma \equiv \cos
\theta \cos \tilde \theta+\sin \theta \sin \tilde \theta \cos
(\varphi -\tilde \varphi)$, $C_{\omega l}$ is a normalization
constant, $\rho_<$ and $\rho_>$ represent the lesser and greater
of $\rho$ and $\tilde \rho$, respectively, and the modes
$p_{\omega l}(\rho)$ and $q_{\omega l}(\rho)$ satisfy the equation
   \bear\label{modeeqn} \left\{ {
   \frac{d^2}{d\rho^2}+\left[\frac{1}{2f}\frac{df}{d\rho}
   +\frac{1}{r^2}\frac{dr^2}{d\rho}\right]\frac{d}{d\rho}
   -\left[\frac{\omega^2}{f}+\frac{l(l+1)}{r^2}+m^2
   +\xi R\right]}\right\}\left\{
   {\begin{array}{l}p_{\omega l}\\
   q_{\omega l}\end{array}}\right\}=0
   \ear
and the Wronskian condition
        \beq\label{wronskian}
        C_{\omega l}\left[p_{\omega l}\frac{dq_{\omega l}}{d\rho}-
        q_{\omega l}\frac{dp_{\omega l}}{d\rho}\right]=\frac{-1}{r^2f^{1/2}}.
        \eeq
Above, it is assumed that the field is in a zero temperature
vacuum state defined with respect to the timelike Killing vector.

By the change of variables
        \beq\label{modes}
        \begin{array}{l}
        \displaystyle
        p_{\omega l}=\frac1{\sqrt {2 r^2 W}}
        \exp \left\{ \int^\rho W f^{-1/2} d\rho \right\}, \\
        \\
        \displaystyle
        q_{\omega l}=\frac1{\sqrt{2 r^2 W}}
        \exp\left\{- \int^\rho W f^{-1/2} d\rho \right\},
        \end{array}
        \eeq
one sees that the Wronskian condition (\ref{wronskian}) is satisfied by
$C_{\omega l}=1$ and the mode equation (\ref{modeeqn})
gives the following equation for $W(\rho)$:
     \beq \label{W2}
     W^2=\omega^2+\frac{f}{r^2} l(l+1)+\frac{f}{4 r^2}+V
     +{ \frac{f'}{8}\frac{(W^2)'}{W^2}
     +\frac{f}{4}\frac{(W^2)''}{W^2}-\frac{5f}{16}\frac{(W^2)'^2}{W^4} }
     ,
     \eeq
where
     \bear
     V&=&f m^2+\left(2\xi-\frac14\right) \frac{f}{r^2}
     +f\left( { \frac{(r^2)''}{2r^2}+\frac{f'(r^2)'}{4fr^2}
      -\frac{(r^2)'^2}{4r^4}} \right)  \nn  &&
    + \xi f \left( {-\frac{f''}{f}-2\frac{(r^2)''}{r^2}
    +\frac{f'^2}{2f^2} +\frac{(r^2)'^2}{2r^4}-\frac{f'(r^2)'}{fr^2}
     } \right).
     \ear
The prime denotes a derivative with respect to $\rho$.

Now one can rewrite expressions (\ref{phi2}),(\ref{Tmnunren})
using expressions (\ref{GE}),(\ref{modes}) and then suppose
$\rho=\tilde \rho, \ \theta=\tilde \theta, \ \phi=\tilde \phi$.
The superficial divergences in the sums over $l$ that appear in
this case can be removed as in Refs. \cite{HC,AHS,CH2,H,A}:
     \beq
     \left< \varphi^2 \right>_{unren}=B_1,
     \eeq
     \bear \label{Ttt}
     \left< T^t_t \right>_{unren}&=&\left[\frac12 g^{t \tilde t}-\frac{\xi}{f}
     -\xi f (g^{t \tilde t})^2-\xi(g^{\rho \tilde t})^2 \right]\frac{f}{r^2}
     \frac{\partial^2 }{\partial \varepsilon^2} B_1
     +\left( 2\xi-\frac12 \right)g^{\rho \tilde \rho} B_2
     +\frac 1{r^2}\left[ \xi f (g^{t \tilde \rho})^2 \right. \nn &&
      \left.+\left(2\xi-\frac12  \right)  \right]B_3
     +\xi\left[ -\frac{f'}{2f}-\frac{ff'}{2}(g^{t \tilde t})^2
     +f\left( -\frac{f'}{2f}-\frac{(r^2)'}{r^2} \right)(g^{t \tilde \rho})^2
     \right]B_4 \nn
     &&+\left[ \xi f\left(- \frac{1}{4r^2}+m^2+\xi R \right)
    (g^{t\tilde \rho})^2
    -\left( 2\xi-\frac12 \right)\frac{1}{4r^2}+\left( 2\xi-\frac12 \right)
    (m^2+\xi R)  \right. \nn && \left.
     +\xi R^t_t  \right] B_1
     +i \xi f'g^{t\tilde t}g^{t\tilde\rho}\sqrt{\frac{f}{r^2}}
      \frac{\partial}{\partial\varepsilon}B_1+i \xi \left[2 g^{t\tilde \rho}
     +2 f g^{t\tilde t} g^{\rho\tilde t} \right] \sqrt{\frac{f}{r^2}}
     \frac{\partial}{\partial\varepsilon}B_4,
     \ear
     \bear \label{Trr}
     \left< T^{\rho}_{\rho} \right>_{unren}&=&\left[\left(2\xi-\frac12\right)
      g^{t \tilde t}+\frac{\xi}{f}+ \xi (g^{\rho \tilde t})^2
     +\frac{\xi}{f}(g^{\rho \tilde \rho})^2 \right] \frac{f}{r^2}
     \frac{\partial^2 }{\partial \varepsilon^2} B_1
     +\frac12 g^{\rho \tilde \rho} B_2 \nn
     &&+\frac 1{r^2}\left[\xi- \xi (g^{\rho \tilde \rho})^2-\frac12
     \right]B_3 +\left[\xi\left( \frac{f'}{2f}+\frac{(r^2)'}{r^2}\right)
     (1+(g^{\rho \tilde \rho})^2)
     +\frac{\xi f'}{2}(g^{\rho \tilde t})^2 \right]B_4 \nn
     &&+\left[ \xi (1+(g^{\rho \tilde \rho})^2) \left( \frac{1}{4r^2}-m^2
    -\xi R \right)
     -\left( 2\xi-\frac12 \right)\frac{1}{4r^2}
    +\left( 2\xi-\frac12 \right)(m^2+\xi R)
     \right. \nn && \left.+\xi R^{\rho}_{\rho}  \right] B_1
     +i \xi \frac{f'}{f}g^{\rho\tilde \rho}g^{t\tilde \rho}
     \sqrt{\frac{f}{r^2}} \frac{\partial}{\partial\varepsilon}B_1
    +i \xi \left[ 2 g^{t\tilde \rho}+2 g^{\rho\tilde \rho} g^{\rho\tilde t}
    \right]
    \sqrt{\frac{f}{r^2}}\frac{\partial}{\partial\varepsilon}B_4,
    \ear
    \bear \label{Tthth}
    \left< T^{\theta}_{\theta} \right>_{unren}&=&\left(2\xi-\frac12\right)
    g^{t \tilde t} \frac{f}{r^2} \frac{\partial^2 }{\partial \varepsilon^2} B_1
    +\left(2\xi-\frac12 \right) g^{\rho \tilde \rho} B_2 +\frac {2\xi}{r^2}B_3
    -\frac{\xi}{2}\frac{(r^2)'}{r^2}B_4 \nn
    &&+\left[-\frac{\xi}{2 r^2}
   +\left( 2\xi-\frac12 \right)(m^2+\xi R)+\xi R^{\theta}_{\theta}  \right] B_1
   +i 2\left(2\xi-\frac12 \right) g^{t\tilde \rho}
   \sqrt{\frac{f}{r^2}}\frac{\partial}{\partial\varepsilon}B_4,
   \ear
where
     \bear\label{B1}
     B_1&=&\frac{1}{4\pi^2}\int \nolimits_{0}^{\infty}du \cos(u \varepsilon)
     \sum \limits_{l=0}^{\infty}\frac1{r^2}
     \left[ \sqrt{\frac{f}{r^2}} \frac{(l+1/2)}{W}-1 \right],
      \ear
     \bear
     B_2&=&\frac{1}{4\pi^2}\int \nolimits_{0}^{\infty}du \cos(u \varepsilon)
     \sum \limits_{l=0}^{\infty} \frac1{r^4}
     \left[-\left(l+ \frac{1}{2} \right) \sqrt{\frac{r^2}{f}}W
    +\left(l+ \frac{1}{2} \right)\frac{{(r^2)'}^2}{4r^2}\sqrt{\frac{f}{r^2}}
    \frac{1}{W}
    \right.\nn
     &&+\left(l+ \frac{1}{2} \right)\frac{{(r^2)}'}{4}\sqrt{\frac{f}{r^2}}
    \frac{{(W^2)}'}{W^3}
    +\left(l+ \frac{1}{2} \right)\frac{r^2}{16}\sqrt{\frac{f}{r^2}}
    \frac{{{(W^2)}'}^2}{W^5}
     +\left(l+ \frac{1}{2} \right)^2
     +\frac{u^2}{2}  \nn
     &&+\frac{r^2}{2f}V
     -\frac{{{(r^2)}'}^2}{4r^2}
    +\frac{r^2 {f}'}{16f}{\left( \frac{f}{r^2} \right)}' \left( \frac{r^2}{f}
    \right)
     -\frac{{(r^2)}'}{4}{\left( \frac{f}{r^2} \right)}' \left( \frac{r^2}{f}
     \right)
    +\frac{r^2}{8}{\left( \frac{f}{r^2} \right)}'' \left( \frac{r^2}{f} \right)
    \nn &&\left.
     -\frac{7r^2}{32}{{\left( \frac{f}{r^2} \right)}'}^{\,2}
     \left( \frac{r^2}{f} \right)^{-2} \right],
     \ear
     \bear
     B_3&=&\frac{1}{4\pi^2}\int \nolimits_{0}^{\infty}du \cos(u \varepsilon)
     \sum \limits_{l=0}^{\infty}\frac1{r^2}
     \left[ \sqrt{\frac{f}{r^2}}\frac{(l+1/2)^3}{W}-\left(l+\frac12 \right)^2
     +\frac{u^2}{2}+\frac{r^4}{8f}\left( \frac{f}{r^2}
    \right)''\right. \nn
     &&\left.+\frac{r^4 {f}'}{16f^2}{\left( \frac{f}{r^2} \right)}'
    -\frac{5r^4}{32f}{{\left( \frac{f}{r^2} \right)}'}^2+\frac{r^2}{2f}V  \right],
    \ear
    \bear\label{B4}
    B_4&=&\frac{1}{4\pi^2}\int \nolimits_{0}^{\infty}du \cos(u \varepsilon)
    \sum \limits_{l=0}^{\infty}\frac1{\left| r \right|^3 }
    \left[ \left( l+\frac12 \right)\frac{{{(r^2)}'}^2}{4r^2} \sqrt{\frac{f}{r^2}}
    \frac{1}{W}-\left( l+\frac12 \right)\frac{\left| r \right| }{4}
      \sqrt{\frac{f}{r^2}}\frac{{(W^2)}'}{W^3}\right.\nn
      &&\left.+\frac{{(r^2)}'}{4\left| r \right|}
      +\frac{\left| r \right| {f}'}{4f}\right],
       \ear
      \beq \label{u}
      \varepsilon=\sqrt{\frac{f}{r^2}}(\tau-\tilde \tau), \quad
      u=w \sqrt{\frac{r^2}{f}}.
      \eeq

\section{WKB approximation for $\langle \varphi^2 \rangle$ and
$\langle T^{\mu}_{\nu} \rangle$ of a very massive field}

Obtaining the exact solution of Eq. (\ref{W2}) is a very
complicated problem. However, it can be solved iteratively if
there is a small parameter of the considered task.

Let us evaluate the order of terms in this equation. If  the
characteristic scale of variation of the gravitational field is
designated as $L$:
       \beq \label{Lm}
       L^{-1}= \max \left \{
       |1/r|, \left| \left[ \ln(fr^2) \right]'  \right|, \
       \left| \left[ \ln(fr^2) \right]''  \right|^{1/2}, \
       \left| \left[ \ln(fr^2) \right]'''  \right|^{1/3}, \
       \dots  \right \},
       \eeq
then in case $\rc=1/m \ll L$, one can consider the quantity
      \beq \label{ewkb}
      \varepsilon_{\mbox{\tiny WKB}}=\frac{\rc} L \ll 1
      \eeq
as a small parameter of the iterative procedure. In this case as
the zeroth-order term of the iterative procedure one can choose
       \beq
       (W^2)_{(0)} = \omega^2+\frac{f}{r^2} l (l+1)+f m^2.
       \eeq
The term of the next order [the iteration procedure demands the
using of $(W^2)_{(0)}$ instead of $W^2$ in the right hand side of
Eq. (\ref{W2}) for calculation of the next order term]
      \beq
      (W^2)_{(2)} =\frac{f}{4 r^2}+V+
      \frac{f'}{8} \frac{ {(W^2)_{(0)}}'}{(W^2)_{(0)}}
      +\frac{f}{4} \frac{{(W^2)_{(0)}}''}{(W^2)_{(0)}}
      -\frac{5f}{16} \frac{{{(W^2)_{(0)}}'}^2}{{(W^2)_{(0)}}^2}
      \eeq
has the order $(\varepsilon_{\mbox{\tiny WKB}})^2 (W^2)_{(0)}$,
i.e., for all values of $w$ and $l$ this term is much less than
the zeroth-order term. For derivation it is convenient to choose
as the zeroth-order term of the iterative procedure the following:
       \beq \label{W0}
       (W^2)_{(0)} = \omega^2+\frac{f}{r^2}
       \left(l+\frac12\right)^2+f m^2,
       \eeq
because the addition to $(W^2)_{(0)}$ some of the addends from
$(W^2)_{(2)}$ does not change the conclusions:
      \beq
      (W^2)_{(0)} \gg (W^2)_{(2)} \gg (W^2)_{\!(4)} \gg \ldots
      \eeq
or
      \beq
       (W^2)_{\!(2)} \sim \varepsilon^{ 2}_{\mbox{\tiny WKB}}
       (W^2)_{(0)}, \quad (W^2)_{\!(4)} \sim
       \varepsilon^{ 4}_{\mbox{\tiny WKB}} (W^2)_{(0)},
       \quad \ldots
      \eeq
and
      \beq
      W^2=(W^2)_{(0)}+(W^2)_{(2)}+(W^2)_{\!(4)}+\ldots .
      \eeq
This approach gives (Ref. \cite{AHS}) the DeWitt-Schwinger
expansions of $\langle \varphi^2 \rangle$ and $\langle
T^{\mu}_{\nu} \rangle$ in terms of powers of $1/(m L)$. Note that
there are the different methods of derivation of the
approximations for $\langle \varphi^2 \rangle$ and $\langle
T^{\mu}_{\nu} \rangle$ in the large mass limit \cite{MatK}.

\section{High-frequency contribution to $\langle \varphi^2 \rangle$
and $\langle T^{\mu}_{\nu} \rangle$}

In the case $\rc \gtrsim L$ or for a massless field a small
parameter does not exist. But we can evaluate the contribution to
$\langle T^{\mu}_{\nu} \rangle$ of the high-frequency modes. For
that it is necessary to impose a lower-limit cutoff $w_0$ on the
integrals over $w$ in expressions (\ref{B1})-(\ref{B4}) ($u \geq
u_0 = w_0 \sqrt{r^2 / f}$). As a small parameter of the iterative
procedure we can choose
     \beq
     \varepsilon_{\mbox{\tiny WKB}}=\frac{\sqrt{f}}{w_0 L}
     =\frac{|r|}{u_0 L}  \ll 1.
     \eeq
This also means [see the definitions of $L$, Eq. (\ref{Lm}), and
$u$, Eq. (\ref{u})]
     \beq
     u_0 \gg \frac{|r|}{L} \geq 1.
     \eeq
The zeroth-order solution of Eq. (\ref{W2}) can be chosen as
follows:
       \beq \label{W02}
       (W^2)_{(0)} = \omega^2+\frac{f}{r^2}
       \left(l+\frac12\right)^2.
       \eeq
Then the second and fourth orders are
      \beq
      (W^2)_{(2)} =V+
      \frac{f'}{8}\frac{{(W^2)_{(0)}}'}{{(W^2)_{(0)}}}
      +\frac{f}{4}\frac{{(W^2)_{(0)}}''}{{(W^2)_{(0)}}}
      -\frac{5f}{16} \frac{{{(W^2)_{(0)}}'}^2}{{(W^2)_{(0)}}^2},
      \eeq
      \bear
      {(W^2)_{\!(4)}}&=&
      \frac{f'}{8}\frac{{(W^2)_{(2)}}'}{{(W^2)_{(2)}}}
      +\frac{f}{4}\frac{{(W^2)_{(2)}}''}{{(W^2)_{(2)}}}
      -\frac{5f}{16} \frac{{{(W^2)_{(2)}}'}^2}{{(W^2)_{(2)}}^2}
       \nn &=& \frac{1}{(W^2)_{(0)}}\left[\frac{f' V'}{8}
      +\frac{f V''}{4}\right]
      +\frac{1}{{(W^2)_{(0)}}^2}
      \left[\left(\frac{f' f''}{64}-\frac{5fV'}{8}+\frac{f f'''}{32}
      -\frac{f' V}{8}\right){(W^2)_{(0)}}'
      \right. \nn && \left.+\left(\frac{f f''}{8}
      +\frac{3 f'^2}{64}-\frac{f V}{4}\right)
      {(W^2)_{(0)}}''
      + \frac{3ff'}{16}{(W^2)_{(0)}}'''
      + \frac{f^2}{16}{(W^2)_{(0)}}''''\right] \nn
      &&+\frac{1}{{(W^2)_{(0)}}^3}\left[
      \left(-\frac{9 f'^2}{128}-\frac{7ff''}{32}
      +\frac{5fV}{8}\right){{(W^2)_{(0)}}'}^2
      -\frac{15ff'}{16}{(W^2)_{(0)}}'
      {(W^2)_{(0)}}'' \right. \nn && \left.
      -\frac{9f^2}{32}{{(W^2)_{(0)}}''}^2
      -\frac{7f^2}{16}{(W^2)_{(0)}}' {(W^2)_{(0)}}'''
      \right]
      +\frac{1}{{(W^2)_{(0)}}^4}\left[
      \frac{27ff'}{32}{{(W^2)_{(0)}}'}^3 \right. \nn && \left.
      +\frac{27f^2}{16} {{(W^2)_{(0)}}'}^2{(W^2)_{(0)}}''
      \right]
      -\frac{1}{{(W^2)_{(0)}}^{5}}\left[
      \frac{135f^2}{128}{{(W^2)_{(0)}}'}^4
      \right].
      \ear
The high-frequency contributions to the quantities $B_1$, $B_2$,
$B_3$, and $B_4$ are obtained by substituting the WKB expansion of
$W^2$ into expressions (\ref{B1})-(\ref{B4}) and imposing a
lower-limit cutoff $u_0$ on the integrals over $u$:
     \bear\label{B1a}
B^{\mbox{\tiny HFC}}_1&=&\frac{1}{4 \pi^2}\left\{
\frac{1}{r^2}S^0_0(\varepsilon, u_0)
-\frac{V}{2f}S^0_1(\varepsilon, u_0)
-\frac{r^2}{16f^2}\left[f'{\left(\frac{f}{r^2}\right)}'
+2f{\left(\frac{f}{r^2}\right)}''\right] S^1_2(\varepsilon, u_0)
\right. \nn && +\frac{5r^4}{32f^2}{{\left(\frac{f}{r^2}
\right)}'}^2 S^2_3(\varepsilon, u_0) +\frac{r^2}{16f^2} \left[
6V^2-f'V'-2f V''\right] S^0_2(\varepsilon, u_0) \nn &&
+\frac{r^4}{128f^3}\left[\left(20Vf' +40fV'-f'f''-2ff'''\right)
{\left(\frac{f}{r^2}\right)}'+ \left(40fV-3{f'}^2 \right. \right.
\nn && \left. \left.-8ff''\right) {\left(\frac{f}{r^2}\right)}''
 -12ff'{\left(\frac{f}{r^2}\right)}'''
-4f^2{\left(\frac{f}{r^2}\right)}''''
 \right]S^1_3(\varepsilon, u_0)
 +\frac{r^6}{512f^4}\left[ \left(21{f'}^2
 \right. \right. \nn && \left. +56ff''-280fV\right)
{{\left(\frac{f}{r^2}\right)}'}^2
+84f^2{{\left(\frac{f}{r^2}\right)}''}^2 +112f^2
{\left(\frac{f}{r^2}\right)}' {\left(\frac{f}{r^2}\right)}''' \nn
&& \left. +252ff'{\left(\frac{f}{r^2}\right)}'
{\left(\frac{f}{r^2}\right)}'' \right]S^2_4(\varepsilon, u_0)
+\frac{r^8}{512f^5}\left[-231ff'{{\left(\frac{f}{r^2}\right)}'}^3
\right. \nn && \left. \left.-
462f^2{{\left(\frac{f}{r^2}\right)}'}^2
{\left(\frac{f}{r^2}\right)}'' \right]S^3_5(\varepsilon, u_0)
+\frac{r^{10}}{2048f^6}
\left[1155f^2{{\left(\frac{f}{r^2}\right)}'}^4 \right]
S^4_6(\varepsilon, u_0)\right\}\nn &&
+O\left(\frac{\varepsilon^2_{\mbox{\tiny WKB}}}{L^2}\right),
     \ear
     \bear\label{B2a}
B^{\mbox{\tiny HFC}}_2&=&\frac{1}{4\pi^2} \left\{ -\frac{1}{r^4}
S^0_{-1}(\varepsilon,u_0) +{\frac {1}{4 f r^6}}\left[{{\left(r^2
\right)}'}^{2}f-2V r^4\right] S^0_0(\varepsilon,u_0) \right. \nn
&& +\frac {1}{16 f^2 r^2}\left[4f\left( r^2 \right)'\left(
\frac{f}{r^2} \right)' -r^2f'\left( \frac{f}{r^2} \right)'-2 f
r^2\left(\frac{f}{r^2}\right)'' \right] S^1_1(\varepsilon,u_0) \nn
&&+\frac{7 r^2}{32 f^2}\left( \frac{f}{r^2} \right)'^{2}
S^2_2(\varepsilon, u_0)+\frac{1}{16f^2r^4}\left[4f r^2 {(r^2)}' V'
-2f {{(r^2)}'}^2 V -r^4 f' V' \right. \nn && \left. +2r^4 V^2 - 2f
r^4 V''\right] S^0_1(\varepsilon, u_0)
 +\frac{1}{128 f^3 r^2} \left[ \left(56 f r^4 V' -48 f r^2 V {(r^2)}'
\right. \right. \nn && \left. -2 f f' {{(r^2)}'}^2 -2 f r^4 f'''
+12 r^4 V f'  -r^4 f' f''  +4 f r^2 f'' {(r^2)}'
\right){\left(\frac{f}{r^2}\right)}'  \nn &&
 +\left(24 f r^4 V  -8 f r^4 f'' -4 f^2 {{(r^2)}'}^2
 +12 f r^2 f' {(r^2)}'  -3 r^4 {f'}^2 \right)
 {\left(\frac{f}{r^2}\right)}'' \nn && \left.
 +\left(8 f^2 r^2 {(r^2)}' -12 f r^4 f' \right)
{\left(\frac{f}{r^2}\right)}'''  -4 f^2 r^4
{\left(\frac{f}{r^2}\right)}'''' \right] S^1_2(\varepsilon, u_0)
\nn && +\frac{1}{512f^4}\left[ \left( 19 r^4 {f'}^2 +20 f^2
{{(r^2)}'}^2 -80f r^2 f' {(r^2)}' -280 f r^4 V \right. \right. \nn
&& \left. +64 f r^4 f'' \right) {{\left( \frac{f}{r^2}
\right)}'}^2 +\left(268 f r^4 f' -160 f^2 r^2 {(r^2)}' \right)
{\left( \frac{f}{r^2} \right)}' {\left( \frac{f}{r^2} \right)}''
\nn && \left. +76 f^2 r^4 {{\left( \frac{f}{r^2} \right)}''}^2
+128 f^2 r^4 {\left( \frac{f}{r^2} \right)}'{\left( \frac{f}{r^2}
\right)}''' \right] S^2_3(\varepsilon, u_0) \nn && +\frac{r^4}{512
f^4 } \left[ \left( 140 f {(r^2)}'  -259 r^2 f' \right) {{\left(
\frac{f}{r^2} \right)}'}^3 -518 f r^2 {{\left( \frac{f}{r^2}
\right)}'}^2 {\left( \frac{f}{r^2} \right)}'' \right]
S^3_4(\varepsilon, u_0) \nn && + \frac{1365 r^8}{2048 f^4 }
{{\left( \frac{f}{r^2} \right)}'}^4 S^4_5(\varepsilon, u_0)
+O\left(\frac{\varepsilon^2_{\mbox{\tiny WKB}}}{L^4}\right),
     \ear
     \bear\label{B3a}
B^{\mbox{\tiny HFC}}_3&=& \frac{1}{4\pi^2} \left\{ {\frac
{1}{r^2}} S^1_0(\varepsilon,u_0) -{\frac {V}{2f}}
S^1_1(\varepsilon,u_0) -\frac{r^2}{16 f^2}\left[f'\left(
\frac{f}{r^2} \right)' +2 f\left( \frac{f}{r^2} \right)''\right]
S^2_2(\varepsilon,u_0) \right.  \nn &&  +{\frac
{5r^4}{32f^2}}{{\left( \frac{f}{r^2} \right)}'}^{2}
S^3_3(\varepsilon,u_0) +\frac{r^2}{16f^2} \left[ 6 V^2 -f' {V}' -2
f {V}'' \right] S^1_2(\varepsilon,u_0) \nn && +\frac{r^4}{128 f^3}
\left[ \left( 40 f V' +20 f' V -2 f f''' -f' f'' \right) {\left(
\frac{f}{r^2} \right)}' +\left( 40 f V -3 {f'}^2 \right. \right.
\nn && \left. \left. -8 f f'' \right) {\left( \frac{f}{r^2}
\right)}'' -12 f f' {\left( \frac{f}{r^2} \right)}''' -4 f^2
{\left( \frac{f}{r^2} \right)}'''' \right] S^2_3(\varepsilon, u_0)
+\frac{r^6}{512 f^4} \left[ \left( 21 {f'}^2  \right. \right. \nn
&& \left. \left. +56 f f'' - 280 f V \right) {{\left(
\frac{f}{r^2} \right)}'}^2 +84 f^2 {{\left( \frac{f}{r^2}
\right)}''}^2 +252 f f' {\left( \frac{f}{r^2} \right)}' {\left(
\frac{f}{r^2} \right)}'' \right. \nn && \left. +112 f^2 {\left(
\frac{f}{r^2} \right)}' {\left( \frac{f}{r^2} \right)}''' \right]
S^3_4(\varepsilon, u_0) -\frac{231 r^8}{512 f^4} \left[ f'
{{\left( \frac{f}{r^2} \right)}'}^3 \right.  \nn && \left. \left.
+2 f {\left( \frac{f}{r^2} \right)}'' {{\left( \frac{f}{r^2}
\right)}'}^2 \right] S^4_5(\varepsilon, u_0) +\frac{1155
r^{10}}{2048 f^4} {{\left( \frac{f}{r^2} \right)}'}^4
S^5_6(\varepsilon, u_0) \right\}
+O\left(\frac{\varepsilon^2_{\mbox{\tiny WKB}}}{L^2}\right),
      \ear
      \bear\label{B4a}
B^{\mbox{\tiny HFC}}_4&=&\frac{1}{4\pi^2} \left\{ -\frac
{\left(r^2 \right)'}{2 r^4} S^0_0(\varepsilon,u_0) -\frac
{1}{4f}{\left(\frac{f}{r^2} \right)}' S^1_1(\varepsilon,u_0)
+\frac{1}{4 f r^2}\left[ {(r^2)}' V -r^2 V' \right]
S^0_1(\varepsilon,u_0) \right. \nn && \frac{1}{32 f^2} \left[
\left( f' {(r^2)}' -r^2 f'' +12 r^2 V \right) {\left(\frac{f}{r^2}
\right)}' +\left( 2 f {(r^2)}' -3 r^2 f' \right)
{\left(\frac{f}{r^2} \right)}'' \right. \nn && \left. -2 f r^2
{\left(\frac{f}{r^2} \right)}''' \right] S^1_2(\varepsilon,u_0)
+\frac{r^2}{64 f^3} \left[ \left( 10 r^2 f' -5 f {(r^2)}' \right)
{{\left(\frac{f}{r^2} \right)}'}^2  \right. \nn && \left. +20 f
r^2 {\left(\frac{f}{r^2} \right)}' {\left(\frac{f}{r^2} \right)}''
\right] S^2_3(\varepsilon,u_0) -\frac{35 r^6}{128 f^3}
{{\left(\frac{f}{r^2} \right)}'}^3 S^3_4(\varepsilon, u_0)
+O\left(\frac{\varepsilon^2_{\mbox{\tiny WKB}}}{L^3}\right).
       \ear
The mode sums and the integrals in these expressions are of the
form
      \beq \label{intsums}
      S^k_n(\varepsilon,u_0)=\int \nolimits_{u_0}^{\infty}du
      \cos (\varepsilon u)
      \sum \limits_{l=0}^{\infty}  \left\{  \frac{\left(l+1/2\right)^{2k+1}}
      {\left[u^2+\left(l+1/2\right)^2\right]^{(2n+1)/2}}-
      \mbox{subtraction terms} \right\},
      \eeq
where $k$ and $n$ are integers, $k \geq 0$ and $n \geq -1$. The
subtraction terms for the sum over $l$  can be obtained by
expanding the function being summed in inverse powers of $l$ and
truncating at $O(l^0)$. This subtraction corresponds to the
removing of the superficial divergences in the sums over $l$
discussed above:
      \bear \label{36}
      S^n_{n-1}(\varepsilon,u_0)&=&\int \nolimits_{u_0}^{\infty}du
      \cos (\varepsilon u)
      \sum \limits_{l=0}^{\infty}  \left\{\frac{\left(l+1/2
      \right)^{2n+1}}{[u^2+\left(l+1/2\right)^2]^{(2n-1)/2}}
      -\left(l+\frac12 \right)^2+(2n-1)\frac{u^2}{2} \right\},
      \ear
      \beq \label{37}
      S^n_n(\varepsilon,u_0)=\int \nolimits_{u_0}^{\infty}du
      \cos (\varepsilon u)
      \sum \limits_{l=0}^{\infty}  \left\{  \frac{\left(l+1/2
      \right)^{2n+1}}{\left[u^2+\left(l+1/2\right)^2
      \right]^{(2n+1)/2}}- 1 \right\}.
      \eeq
For the other quantities $S^k_n(\varepsilon,u_0)$ there are no
subtraction terms. The details of the calculations of
$S^k_n(\varepsilon,u_0)$ in the limit $\varepsilon \rightarrow 0$
are discussed in Appendix A:
     \bear \label{start}
     S^{0}_{-1}(\varepsilon, u_0)&=&-\frac{2}{\varepsilon^4}
     -\frac{1}{24 \varepsilon^2}+\frac{u_0^4}{12}
     -\frac{u_0^2}{48}+\frac{7}{1920} \left( C+\frac12\ln\left|
     \varepsilon^2 u_0^2\right|\right)-\frac{31}{129024}
     \frac1{u_0^2}\nn && +O\left( \frac{1}{u_0^4} \right)
     +O\left(\varepsilon^2 \ln\left|\varepsilon \right|\right),
     \ear
     \beq\
     S^1_0(\varepsilon, u_0)=\frac{4}{ \varepsilon^4}
     -\frac{u_0^4}{6}+ \frac{7}{960} \left( C+\frac12\ln
     \left| \varepsilon^2 u_0^2
     \right|\right)-\frac{31}{32256} \frac{1}{u_0^2}
     +O\left( \frac{1}{u_0^4} \right)
     +O\left( \varepsilon^2 \ln \left| \varepsilon \right| \right),
     \eeq
     \bear
     S^{0}_{0}(\varepsilon, u_0)&=&\frac{1}{\varepsilon^2}
     +\frac{u_0^2}{2}-\frac{1}{24}\left( C+\frac12\ln\left|
     \varepsilon^2 u_0^2 \right|\right)
     +\frac{7}{3840} \frac{1}{u_0^2}+\varepsilon^2 \left[
     -\frac{u_0^4}{8} +\frac{u_0^2}{96} -\frac{7}{2560}
     \right.\nn && \left. +\frac{7}{3840} \left(C+\frac12\ln
     \left| \varepsilon^2 u_0^2 \right|\right)
     -\frac{31}{86016} \frac{1}{u_0^2}\right]
     +O\left( \frac{1}{u_0^4} \right)
     +O\left( \frac{\varepsilon^2}{u_0^4} \right)
     +O\left( \varepsilon^4 \right),
     \ear
     \bear
     S^{1}_{1}(\varepsilon, u_0)&=&\frac{2}{\varepsilon^2}
     +u_0^2-\frac{7}{1920} \frac{1}{u_0^2}
     +O\left( \frac{1}{u_0^4} \right)
     +O\left(\varepsilon^2 \ln\left|\varepsilon\right|\right),
     \ear
     \bear
     S^{n}_{n}(\varepsilon, u_0)&=&\frac{(2n)!!}{(2n-1!!)}
     \frac{1}{\varepsilon^2}+\frac{(2n)!!}{(2n-1!!)}
     \frac{u_0^2}{2} +O\left( \frac{1}{u_0^4} \right)
     +O\left( \varepsilon^2 \ln \left| \varepsilon \right|
     \right) \hspace{2cm} (n\geq 2),
     \ear
     \bear
     S^{0}_{1}(\varepsilon, u_0)&=&-\left( C+\frac12\ln\left|
     \varepsilon^2 u_0^2 \right|\right)+
     \frac{1}{48}\frac{1}{u_0^2}+\varepsilon^2 \left[
     \frac{u_0^2}{4}+\frac{1}{48}
     \left(C+\frac12\ln \left| \varepsilon^2 u_0^2 \right|\right)
     -\frac{1}{32}\right. \nn && \left. -\frac{7}{2560}
     \frac{1}{u_0^2}\right]+O\left( \frac{1}{u_0^4} \right)
     +O\left( \frac{\varepsilon^2}{u_0^4} \right)
     +O\left( \varepsilon^4 \right),
     \ear
     \bear
     S^{1}_{2}(\varepsilon, u_0)&=&-\frac23\left(
     C+\frac12\ln\left|\varepsilon^2 u_0^2 \right|\right)
     +\varepsilon^2 \left[\frac{u_0^2}{6}+\frac{7}{3840}
     \frac{1}{u_0^2}\right]
     +O\left( \frac{1}{u_0^4} \right)
     +O\left( \frac{\varepsilon^2}{u_0^4} \right)
     +O\left( \varepsilon^4 \right),
     \ear
     \bear
     S^{n}_{n+1}(\varepsilon, u_0)&=&\frac{(2n)!!}{(2n+1)!!}
     \left[-\left( C+\frac12\ln\left|\varepsilon^2 u_0^2
     \right|\right)+\varepsilon^2 \frac{u_0^2}{4}\right]
     +O\left( \frac{1}{u_0^4} \right)
     +O\left( \frac{\varepsilon^2}{u_0^4} \right) \nn &&
     +O\left( \varepsilon^4 \right) \hspace{9cm} (n\geq 2),
     \ear
     \bear
     S^{0}_{2}(\varepsilon, u_0)&=&\frac{1}{6 u_0^2}
     +\varepsilon^2 \left[-\frac14+\frac16 \left( C+\frac12
     \ln\left|\varepsilon^2 u_0^2 \right|\right)
     -\frac{1}{96 u_0^2}\right]
     +O\left( \frac{1}{u_0^4} \right)
     +O\left( \frac{\varepsilon^2}{u_0^4} \right) \nn &&
     +O\left( \varepsilon^4 \right) \hspace{9cm} (n\geq 2),
     \ear
     \bear
     S^{n}_{n+2}(\varepsilon, u_0)&=&\frac{(2n)!!}{(2n+3)!!}
     \left\{\frac{1}{2 u_0^2}+\varepsilon^2 \left[-\frac34
     +\frac12 \left( C+\frac12 \ln\left|\varepsilon^2 u_0^2
     \right|\right)\right]\right\}
     +O\left( \frac{1}{u_0^4} \right)\nn &&
     +O\left( \frac{\varepsilon^2}{u_0^4} \right)
     +O\left( \varepsilon^4 \right) \hspace{7cm} (n\geq 1),
     \ear
     \bear
     S^{n}_{n+3}(\varepsilon, u_0)&=&-\frac{(2n)!!}{(2n+5)!!}
     \frac{3 \varepsilon^2}{4 u_0^2}
     +O\left( \frac{1}{u_0^4} \right)
     +O\left( \frac{\varepsilon^2}{u_0^4} \right)
     +O\left( \varepsilon^4 \right) \hspace{2cm} (n\geq 0),
     \ear
     \bear\label{finish}
     S^{k}_{n}(\varepsilon, u_0)&=&O\left( \frac{1}{u_0^4}
     \right)
     +O\left( \frac{\varepsilon^2}{u_0^4} \right)
     +O\left( \varepsilon^4 \right) \hspace{3.5cm} (k\geq 0,
     \ n\geq k+4).
     \ear
The substitution of these expressions into Eqs.
(\ref{B1a})-(\ref{B4a}) and then into Eqs.
(\ref{Ttt})-(\ref{Tthth}) gives $\langle \varphi^2
\rangle^{\mbox{\tiny HFC}}_{unren}$ and $\left< T^{\mu}_{\nu}
\right>^{\mbox{\tiny HFC}}_{unren}$ - the high-frequency
contributions to $\langle \varphi^2 \rangle_{unren}$ and $\left<
T^{\mu}_{\nu} \right>_{unren}$. Let us note that the expansions of
$\langle \varphi^2 \rangle^{\mbox{\tiny HFC}}_{unren}$ and $\left<
T^{\mu}_{\nu} \right>^{\mbox{\tiny HFC}}_{unren}$ in terms of
powers of $u_0$ correspond to the DeWitt-Schwinger expansions of
$\langle \varphi^2 \rangle_{unren}$ and $\left< T^{\mu}_{\nu}
\right>_{unren}$ in terms of powers of $mL$.

The renormalizations of $\langle \varphi^2 \rangle$ and $\left<
T^{\mu}_{\nu} \right>$ are achieved by subtracting the
renormalization counterterms from $\langle \varphi^2
\rangle_{unren}$ and $\left< T^{\mu}_{\nu} \right>_{unren}$ and
then letting $\tilde \tau \rightarrow \tau$:
      \beq
      \left< \varphi^2 \right>_{ren}=
      \lim_{\tilde \tau \rightarrow \tau}
      \left[\left< \varphi^2 \right>_{unren}-
      \left< \varphi^2 \right>_{\mbox{\tiny DS}}\right],
      \eeq
      \beq
      \left< T^{\mu}_{\nu} \right>_{ren}=
      \lim_{\tilde \tau \rightarrow \tau}
      \left[\left< T^{\mu}_{\nu} \right>_{unren}-
      \left< T^{\mu}_{\nu} \right>_{\mbox{\tiny DS}}\right],
      \eeq
where
      \bear
      \phisq_{\mbox{\tiny DS}} &=& \frac1{8\pi^2\sigma}+\frac1{8\pi^2}
      \left[m^2+\left(\xi-\frac16\right)R\right]
      \left[C+\frac12\ln\left( \frac{\mds^2|\sigma|}{2}
      \right)\right] \nn && -\frac{m^2}{16\pi^2}
      +\frac1{96\pi^2}R_{\alpha\beta}
      \frac{\sigma^\alpha\sigma^\beta}{\sigma},
      \ear
$C$ is Euler's constant, $R_{\alpha\beta}$ is the Ricci tensor,
$R=R^{\alpha}_{\alpha}$, $\sigma$ is one-half the square of the
distance between the points $x$ and $\tilde x$ along the shortest
geodesic connecting them, $\sigma^{\mu}$ is the covariant
derivative of $\sigma$, and the constant $\mds$ is equal to the
mass $m$ of the field for a massive scalar field. For a massless
field $\mds$ is an arbitrary parameter due to the infrared cutoff
in $\left< \varphi^2\right>_{\mbox{\tiny DS}}$. A particular
choice of the value of $\mds$ corresponds to a finite
renormalization of the coefficients of terms in the gravitational
Lagrangian and must be fixed by experiment or observation. The
expression for $\left<T^{\mu}_{\nu} \right>_{\mbox{\tiny DS}}$ is
displayed in Ref. \cite{Chris} and is too long to be displayed
here.

For the metric (\ref{metric}) the calculations of quantities
$\sigma^{\mu}$ and $g^{\mu \tilde \nu}$ are similar to those in
Ref. \cite{AHS}:
      \bear
      \sigma^t&=&(t-\tilde t)+\frac{f'^2}{24f}(t-\tilde t)^3
      -\frac{1}{120}\left(\frac{f'^{\,4}}{8f^2}-
      \frac{3}{8}\frac{f'^{\,2} f''}{f}
      \right)(t-\tilde t)^5
      +O\left((t-\tilde t)^7\right),\nn
      \sigma^{\rho}&=&-\frac{f'}{4}(t-\tilde t)^2
      -\frac{f' f''}{96}(t-\tilde t)^4
      +O\left((t-\tilde t)^6\right),\nn
      \sigma^{\theta}&=&\sigma^{\phi}=0, \nn
      \sigma&=&\frac12 g_{\mu \nu}\sigma^{\mu}\sigma^{\nu},
      \ear
      \bear
      g^{t \tilde t}&=&-\frac{g^{\rho \tilde \rho}}{f}=
      -\frac1f-\frac{f'^2}{8f^2}(t-\tilde t)^2
      +\left(\frac{f'^{\,4}}{384f^3}-\frac{f'^{\,2}f''}{96f^2}
      \right)(t-\tilde t)^4
      +O\left((t-\tilde t)^6\right),\nn
      g^{t \tilde \rho}&=&-g^{\rho \tilde t}=
      -\frac{f'}{2f}(t-\tilde t)
      -\left(\frac{f'^{\,3}}{96 f^2}+\frac{f'f''}{48f}
      \right)(t-\tilde t)^3
      +O\left((t-\tilde t)^5\right).
      \ear
The high-frequency contributions to $\left< \varphi^2
\right>_{unren}$ and $\left< T^{\mu}_{\nu} \right>_{unren}$
contain all the ultraviolet divergences and can also be
renormalized. This procedure gives the second-order WKB
approximation for $\left< \varphi^2 \right>_{ren}$ and the
fourth-order one for components of $\left< T^{\mu}_{\nu}
\right>_{ren}$ that contain the contributions of high-frequency
modes only ($w > w_0$):
     \beq \label{res1}
     \left< \varphi^2 \right>_{\mbox{\tiny WKB}}=
     \lim_{\tilde \tau \rightarrow \tau}
     \left[\left< \varphi^2 \right>^{\mbox{\tiny HFC}}_{unren}-
     \left< \varphi^2 \right>_{\mbox{\tiny DS}}\right]=
     \left( \varphi^2 \right)_{\!(0)}
     +\left( \varphi^2 \right)_{\!(2)}
     +O\left(\frac{1}{{u_0}^2 L^2}\right),
     \eeq
     \beq  \label{res2}
     \left< T^{\mu}_{\nu} \right>_{\mbox{\tiny WKB}}=
     \lim_{\tilde \tau \rightarrow \tau}
     \left[\left< T^{\mu}_{\nu} \right>^{\mbox{\tiny HFC}}_{unren}-
     \left< T^{\mu}_{\nu} \right>_{\mbox{\tiny DS}}\right]=
     \left( T^{\mu}_{\nu}\right)_{(0)}
     +\left( T^{\mu}_{\nu} \right)_{(2)}
     +\left(T^{\mu}_{\nu} \right)_{\!(4)}
     +O\left(\frac{1}{{u_0}^2 L^4}\right),
     \eeq
where $\left( \varphi^2 \right)_{\!(0)}$, $\left( \varphi^2
\right)_{\!(2)}$, and the nontrivial components  of $\left<
T^{\mu}_{\nu} \right>_{\mbox{\tiny WKB}}$ of zeroth and second WKB
orders have the form
     \beq
     \left( \varphi^2 \right)_{\!(0)}=\frac{{u_0}^2}{8 \pi^2 r^2},
     \eeq
      \bear
      \left(\varphi^2 \right)_{\!(2)}&=& \frac{m^2}{16 \pi^2}
      +\frac{1}{16 \pi^2}\left[m^2+\left(\xi-\frac{1}{6} \right)R
      \right]\ln\left|\frac{4 u_0^2}{\mds^2 r^2 } \right|
      -\frac{{f'}^2}{96 \pi^2 f^2} +\frac{f''}{96 \pi^2 f^2} +\frac{f' {(r^2)}'}{96
      \pi^2 f r^2},
      \ear
     \bear \label{T 0_tt}
     \left( T^{t}_{t} \right)_{(0)}=
     \frac{{u_0}^4}{16 \pi^2 r^4},
     \ear
     \bear \label{T 0_rr}
     \left( T^{\rho}_{\rho} \right)_{(0)}=
     -\frac{{u_0}^4}{48 \pi^2 r^4},
     \ear
     \bear \label{T 0_thth}
     \left( T^{\theta}_{\theta} \right)_{(0)}=
     \left( T^{\varphi }_{\varphi} \right)_{(0)}=
     -\frac{{u_0}^4}{48 \pi^2 r^4},
     \ear
     \bear \label{T 2_tt}
     \left( T^{t}_{t} \right)_{(2)}=
     \frac{{u_0}^2}{4 \pi^2 r^2}\left[
     \left(\xi-\frac16\right)
     \left(\frac{1}{2r^2}+\frac{5 {{f}'}^2}{4 f^2}
     -\frac{f' {(r^2)}'}{f r^2}+\frac{{{(r^2)}'}^2}{8 r^4}
     -\frac{{f}''}{f}-\frac{{(r^2)}''}{2 r^2} \right)
     +\frac{m^2}{4}\right],
     \ear
     \bear \label{T 2_rr}
     \left( T^{\rho}_{\rho} \right)_{(2)}=
     \frac{{u_0}^2}{4 \pi^2 r^2}\left[
     \left(\xi-\frac16\right)
     \left(-\frac{1}{2r^2}-\frac{{{f}'}^2}{4 f^2}
     -\frac{f' {(r^2)}'}{4 f r^2}+\frac{{{(r^2)}'}^2}{8 r^4}
     \right)-\frac{m^2}{4}\right],
     \ear
     \bear \label{T 2_thth}
     \left( T^{\theta}_{\theta} \right)_{(2)}&=&
     \left( T^{\varphi }_{\varphi} \right)_{(2)}=
     \frac{{u_0}^2}{4 \pi^2 r^2}\left[
     \left(\xi-\frac16\right)
     \left(\frac{5 {{f}'}^2}{8 f^2}
     -\frac{f' {(r^2)}'}{8 f r^2}-\frac{{{(r^2)}'}^2}{8 r^4}
     -\frac{{f}''}{4 f}+\frac{{(r^2)}''}{4 r^2} \right)
     -\frac{m^2}{4}\right],
     \ear
     \beq
     u_0=w_0 \sqrt{r^2/f}\gg 1.
     \eeq
The quantities of the second WKB order $\left( \varphi^2
\right)_{\!(2)}$ and the fourth WKB order $\left(
T^{\mu}_{\nu}\right)_{\!(4)}$ are equivalent to the analytical
approximations $\left< \varphi^2 \right>_{analytic}$ and $\left<
T^{\mu}_{\nu}\right>_{analytic}$ of Anderson, Hiscock, and Samuel
\cite{AHS} (the constants $m_{\mbox{\tiny DS}}$ and $w_0$ are
equal to $\mu$ and $\lambda$ in their expressions for $\left<
\varphi^2 \right>_{analytic}$ and $\left<
T^{\mu}_{\nu}\right>_{analytic}$, respectively). In the
coordinates (\ref{metric}) the expressions for $\left(
T^{\mu}_{\nu}\right)_{\!(4)}$ are given in Appendix B.

The quantity $\left< T^{\mu}_{\nu} \right>_{\mbox{\tiny WKB}}$ is
conserved and has the trace
     \bear
     \left< T^{\mu}_{\mu} \right>_{\mbox{\tiny WKB}}=
     \frac{{u_0}^2}{4 \pi^2 r^2}\left[-\frac{m^2}{2}
     +\left(\xi-\frac16\right)
     \left(\frac{9 {{f}'}^2}{4 f^2}
     -\frac{3 f' {(r^2)}'}{2 f r^2}-\frac{3{f}''}{2 f}
     \right)\right]+\left(T^{\mu}_{\mu} \right)_{\!(4)}.
      \ear
For a conformally invariant field this trace is equal to the trace
anomaly.


\section{Low-frequency contribution to $\langle \varphi^2 \rangle$
and $\langle T^{\mu}_{\nu} \rangle$}


The behavior of low-frequency modes is determined by the boundary
conditions and the topological structure of the spacetime. If the
spacetime is asymptotically flat and the characteristic scale of
the gravitational field inhomogeneity $\lambda$ is much less than
the parameter $\sqrt{f}/\omega_0$,
     \beq
     \frac{\lambda }{\sqrt{f}/\omega_0} \ll 1,
     \eeq
the low-frequency contributions to $\langle \varphi^2 \rangle$ and
$\langle T^{\mu}_{\nu} \rangle$ can be expanded in terms of powers
of this small parameter. In the following the zeroth term of this
expansion will be used for approximation of the low-frequency
contributions to $\langle \varphi^2 \rangle$ and $\langle
T^{\mu}_{\nu} \rangle$. This means that we choose the long-wave
modes approximately coincident with long-wave modes of the
Minkowski vacuum (in generally accepted terminology this
corresponds to the choice of the Boulware quantum state). For
these modes [$ds^2=dT^2+dx^2+dy^2+dz^2 = dT^2 +dr^2 +r^2(d\theta^2
+\sin^2\theta\, d\varphi^2) $]
      \bear
      G_E(x, \tilde x)&=&\frac{1}{(2\pi)^4}\int d \Omega
      d^3p \frac{\exp\left(i \Omega \Delta T
      +i p_{\alpha} \Delta x^{\alpha}\right)}
      {\left(\Omega^2 +p_x^2 +p_y^2 +p_z^2+m^2\right)}
      \nn &=&\frac{1}{4 \pi^3} \int d\Omega e^{i \Omega
      \Delta T} \int^{\infty}_{0} d p \frac{p \sin(p
      \Delta r)}{\Delta r \left(\Omega^2 +p^2 +m^2
      \right)} \nn &=& \frac{1}{8 \pi^2} \int d\Omega e^{i \Omega
      \Delta T} \frac{\exp \left(-\Delta r
      \sqrt{\Omega^2 +m^2} \right)}{\Delta r} \nn &=&
      \frac{1}{8 \pi^2} \int d\Omega e^{i \Omega
      \Delta T}\left[\frac{1}{\Delta r}-
      \sqrt{\Omega^2+m^2}+O(\Delta r) \right].
      \ear
The first addend in the subintegral function must be removed
because it gives a superficial divergence similar to that
discussed above. Then the low-frequency contribution to $\langle
\varphi^2 \rangle$ is
      \bear
      \langle \varphi^2 \rangle_{\mbox{\tiny LFC}}&=&
      \lim_{\Delta \tau \rightarrow 0}\left\{
      -\frac{1}{8 \pi^2} \int_{-\Omega_0}^{\Omega_0}
      d\Omega e^{i \Omega \Delta \tau}
      \sqrt{\Omega^2+m^2}\right\}\nn &=&  -\frac{1}{8 \pi^2}
      \left(\Omega_0 \sqrt{{\Omega_0}^2+m^2}
      +m^2 \ln \left|\frac{\Omega_0+\sqrt{{\Omega_0}^2 +m^2}}{m}
      \right| \ \right).
      \ear
In the case of a massless field or if $\Omega_0 \gg m$ this
expression can be rewritten as
      \bear
      \langle \varphi^2 \rangle_{\mbox{\tiny LFC}}
      &=& -\frac{{\Omega_0}^2}{8 \pi^2 }
      -\frac{m^2}{16 \pi^2} -\frac{m^2}{16 \pi^2} \ln \left|
      \frac{4 {\Omega_0}^2}{m^2}\right| +O\left(\frac{m^4}{{\Omega_0}^2}
      \right).
      \ear
If we take into account $\Omega_0 =\omega_0/\sqrt{f}$ then
      \bear \label{phiren}
      \langle \varphi^2 \rangle_{ren}&=&
      \lim_{\tilde \tau \rightarrow \tau}
     \left[\langle \varphi^2 \rangle_{unren}-
      \langle \varphi^2 \rangle_{\mbox{\tiny DS}}\right]
      =\lim_{\tilde \tau \rightarrow \tau}
     \left[\langle \varphi^2 \rangle^{\mbox{\tiny HFC}}_{unren}
      -\langle \varphi^2 \rangle_{\mbox{\tiny DS}}\right]
      +\langle \varphi^2 \rangle_{\mbox{\tiny LFC}}=
      \langle \varphi^2 \rangle_{\mbox{\tiny WKB}}
      + \langle \varphi^2 \rangle_{\mbox{\tiny LFC}}
      \nn &=& \frac{R}{16 \pi^2} \left(\xi-\frac16 \right)\ln
      \left| \frac{4 {u_0}^2}{\mds^2 r^2} \right| -\frac{{f'}^2}
      {96 \pi^2 f^2} +\frac{f''}{96 \pi^2f} +\frac{f' {(r^2)}'}
      {96 \pi^2 f r^2} +O\left(\frac{1}{L^2 {u_0}^2} \right).
      \ear
The corresponding expressions for $\left< T^{\mu}_{\mu}
\right>_{\mbox{\tiny LFC}}$ are
       \bear
       \left< T^{t}_{t} \right>_{\mbox{\tiny LFC}}
       &=&-\frac{\Omega_0^4}{16
       \pi^2}-\frac{m^2 \Omega_0^2}{16 \pi^2}-\frac{m^4}{128 \pi^2}
       +\frac{m^4}{64 \pi^2} \ln \left| \frac{4 \Omega_0^2}{m^2}
       \right| +O\left(\frac{m^6}{\Omega_0^2}\right),
       \ear
       \bear
       \left< T^{r}_{r} \right>_{\mbox{\tiny LFC}}
       &=&\left< T^{\theta}_{\theta}\right>_{\mbox{\tiny LFC}}
       =\left< T^{\varphi}_{\varphi} \right>_{\mbox{\tiny LFC}}=
       \frac{\Omega_0^4}{48\pi^2}+\frac{m^2 \Omega_0^2}{16 \pi^2}
       +\frac{3 m^4}{128 \pi^2}+\frac{m^4}{64 \pi^2} \ln \left|
       \frac{4 \Omega_0^2}{m^2}\right|
       +O\left(\frac{m^6}{\Omega_0^2}\right).
       \ear
The analytical approximation for $\left< T^{\mu}_{\nu}
\right>_{ren}$ in the case of asymptotically flat spacetimes is
      \bear
      \left< T^{\mu}_{\nu} \right>_{ren}&=&
      \lim_{\tilde \tau \rightarrow \tau}
     \left[\left< T^{\mu}_{\nu} \right>_{unren}-
      \left< T^{\mu}_{\nu} \right>_{\mbox{\tiny DS}}\right]
      =\lim_{\tilde \tau \rightarrow \tau}
     \left[\left< T^{\mu}_{\nu} \right>^{\mbox{\tiny HFC}}_{unren}
      -\left< T^{\mu}_{\nu} \right>_{\mbox{\tiny DS}}\right]
      +\left< T^{\mu}_{\nu} \right>_{\mbox{\tiny LFC}} \nn &=&
      \left< T^{\mu}_{\nu} \right>_{\mbox{\tiny WKB}}
      +\left< T^{\mu}_{\nu} \right>_{\mbox{\tiny LFC}},
      \ear
     \bear \label{T00}
     \left< T^{t}_{t} \right>_{ren}&=&
     \frac{{u_0}^2}{4 \pi^2 r^2}\left(\xi-\frac16\right)
     \left(\frac{1}{2r^2}+\frac{5 {{f}'}^2}{4 f^2}
     -\frac{f' {(r^2)}'}{f r^2}+\frac{{{(r^2)}'}^2}{8 r^4}
     -\frac{{f}''}{f}
     -\frac{{(r^2)}''}{2 r^2} \right) \nn &&
     -\frac{m^4}{128 \pi^2}
     +\frac{m^4}{64 \pi^2} \ln \left|\frac{4 u_0^2}{m^2 r^2}
     \right|+\left(T^t_t \right)_{(4)}
     +O\left(\frac{1}{u_0^2 L^4} \right)
     +O\left(\frac{\lambda u_0}{r L^4} \right),
     \ear
     \bear \label{T11}
     \left< T^{\rho}_{\rho} \right>_{ren}&=&
     \frac{{u_0}^2}{4 \pi^2 r^2}\left(\xi-\frac16\right)
     \left(-\frac{1}{2r^2}-\frac{{{f}'}^2}{4 f^2}
     -\frac{f' {(r^2)}'}{4 f r^2}+\frac{{{(r^2)}'}^2}{8 r^4}
     \right)  \nn && +\frac{3 m^4}{128 \pi^2}
     +\frac{m^4}{64 \pi^2} \ln \left|\frac{4 u_0^2}{m^2 r^2}
     \right|+\left(T^{\rho}_{\rho} \right)_{(4)}
     +O\left(\frac{1}{u_0^2 L^4} \right)
     +O\left(\frac{\lambda u_0}{r L^4} \right),
     \ear
     \bear \label{T22}
     \left< T^{\theta}_{\theta} \right>_{ren}&=&
     \left< T^{\varphi }_{\varphi} \right>_{ren}=
     \frac{{u_0}^2}{4 \pi^2 r^2} \left(\xi-\frac16\right)
     \left(\frac{{(r^2)}''}{4 r^2}-\frac{{f}''}{4 f}
     +\frac{5 {{f}'}^2}{8 f^2}-\frac{{{(r^2)}'}^2}{8 r^4}
     -\frac{f' {(r^2)}'}{8 f r^2}\right) \nn &&
     +\frac{3 m^4}{128 \pi^2}
     +\frac{m^4}{64 \pi^2} \ln \left|\frac{4 u_0^2}{m^2 r^2}
     \right|+\left(T^{\theta}_{\theta} \right)_{(4)}
     +O\left(\frac{1}{u_0^2 L^4} \right)
     +O\left(\frac{\lambda u_0}{r L^4} \right).
     \ear
This tensor is conserved,
     \beq
     \left< T^{\mu}_{\nu}\right>_{ren;\,\mu}=
     O\left(\frac{1}{u_0^2 L^5} \right)
     +O\left(\frac{\lambda u_0}{r L^5}\right),
     \eeq
and has the trace
     \bear
     \left< T^{\mu}_{\mu} \right>_{ren}&=&
     \frac{{u_0}^2}{4 \pi^2 r^2}\left(\xi-\frac16\right)
     \left(\frac{9 {{f}'}^2}{4 f^2}
     -\frac{3 f' {(r^2)}'}{2 f r^2}-\frac{3{f}''}{2 f}
     \right)-\frac{m^4}{16 \pi^2}+\frac{m^4}{16 \pi^2}
     \ln\left|\frac{4u_0^2}{m^2 r^2}\right|
     \nn &&+\left(T^{\mu}_{\mu} \right)_{\!(4)}
     +O\left(\frac{1}{u_0^2 L^4}\right)
     +O\left(\frac{\lambda u_0}{r L^4} \right).
      \ear
For a conformally invariant scalar field this trace is also equal
to the trace anomaly.


\section{Conclusion}


In this paper, analytical approximations for $\langle \varphi^2
\rangle_{ren}$ and $\left< T^{\mu}_{\nu} \right>_{ren}$ of
quantized scalar fields in static spherically symmetric
asymptotically flat spacetimes have been obtained. The field is
assumed to be in a zero temperature vacuum state, with mass $m
\lsim 1/L$ [$L(\rho)$ is the characteristic scale of variation of
the gravitational field at the considered point] and with an
arbitrary coupling $\xi$ to the scalar curvature.

The necessary conditions for the validity of the analytical
approximations (\ref{phiren}),(\ref{T00})-(\ref{T22}) are
       \beq \label{condition}
       \lambda \ll \frac{\sqrt{f(\rho)}}{w_0} \ll L(\rho),
       \eeq
where $\lambda$ is the characteristic scale of the gravitational
field inhomogeneity in asymptotically flat spacetime and $w_0$ is
the constant of the WKB expansion.

If we consider a spherical body with radius $r_0 > r_g$ ($r_g$ is
the gravitational radius of this body, $\lambda \sim r_0$) then
the metric spacetime outside this body is
\begin{equation}
ds^2=-\left(1-\frac{r_g}{r(\rho)}\right)dt^2 +d\rho^2 +r(\rho)^2
(d\theta^2+\sin^2\theta\, d\varphi^2),
\end{equation}
where $r(\rho)$ is the inverse function to the function
     \bear
     \rho(r)&=&\rho_0+r_g
     \left[\sqrt{\frac{r}{r_g}\left(\frac{r}{r_g}-1
     \right)}-\sqrt{\frac{r_0}{r_g}\left(\frac{r_0}{r_g}-1 \right)}
     \right. \nn && \left. +\frac12 \ln \left|\frac{\left(\sqrt{r/r_g
     -1} +\sqrt{r/r_g}\right)\left(\sqrt{r_0/r_g -1}
     -\sqrt{r_0/r_g}\right)}{\left(\sqrt{r/r_g -1}
     -\sqrt{r/r_g}\right)\left(\sqrt{r_0/r_g -1}
     +\sqrt{r_0/r_g}\right)}\right| \right].
     \ear
When $r(\rho) \gg r_0$
     \begin{equation} f(\rho) \sim 1, \quad L(\rho)
     \sim r(\rho),
     \end{equation}
and the conditions (\ref{condition}) can be satisfied by the
choice of $w_0$,
\begin{equation}
r_0 \ll w_0 \ll r(\rho).
\end{equation}
This means that, far from the body where $r(\rho) \gg r_0$, the
approximations (\ref{phiren}),(\ref{T00})-(\ref{T22}) are valid.

The presence in expressions (\ref{phiren}),(\ref{T00})-(\ref{T22})
the arbitrary parameter $u_0=w_0 r/\sqrt{f}$ is a generic feature
of local approximation schemes \cite{AHS,GAC,A,PS,P}. For a
conformally invariant field this parameter can be absorbed into
the definition of the constant $\mds$.


When the massless conformally coupled scalar field is in a zero
temperature vacuum state and propagating in static spherically
symmetric asymptotically flat spacetimes the expressions
(\ref{phiren}),(\ref{T00})-(\ref{T22}) are equivalent to the
approximations of  Page, Brown, and Ottewill \cite{Page,BO,BOP},
Zannias \cite{Z}, and Frolov and Zel'nikov \cite{FZ} (for the
particular choice of the arbitrary parameters in their
expressions), and the analytical approximation of Anderson,
Hiscock, and Samuel \cite{AHS}. Let us note that in this case the
low-frequency contributions to $\langle \varphi^2 \rangle_{ren}$
and $\left< T^{\mu}_{\nu} \right>_{ren}$ are equivalent to the
low-frequency contributions which are given by the procedure of
\cite{AHS} for modes with $w<w_0$. This means that using such
calculations in \cite{BFNFSZ,FSZ,BFNS} is correct in the case of a
conformally invariant field. Let us note also that the more exact
procedure \cite{Sushkov3} gives different result.

\section*{Acknowledgments}

I would like to thank S. V. Sushkov and N. R. Khusnutdinov for
helpful discussions. This work was supported by the Russian
Foundation for Basic Research Grant No. 02-02-17177 and by the
SRPED Foundation of Tatarstan Republic Grant No. 06-6.5-110.

\section*{Appendix A}

\setcounter{equation}{0}
\renewcommand{\theequation}{A\arabic{equation}}
To calculate the quantities $S^k_n(\varepsilon, u_0)$ it is
necessary to compute the various sums over $l$. We start from the
sum in expression (\ref{36}):
      \bear \label{sun} S(u, n)=
      \sum \limits_{l=0}^{\infty}  \left\{\frac{\left(l+\frac12
      \right)^{2n+1}}{[u^2+\left(l+1/2\right)^2]^{(2n-1)/2}}
      -\left(l+\frac12 \right)^2+(2n-1)\frac{u^2}{2} \right\},
       \quad n \geq 0.
      \ear
For calculation of this sum we can use the Abel-Plana method
\cite{Evg,Sushkov2,P}
      \beq
      \sum \limits_{l=0}^{\infty} F( l+1/2)
      =\int \nolimits_{q}^{\infty}F(x)dx
      +\int \nolimits_{q-i\infty}^{q}\frac{F(z)}{1+e^{i2\pi z}}dz
      -\int \nolimits_{q}^{q+i\infty}\frac{F(z)}{1+e^{-i2\pi z}}dz,
      \eeq
where $-1/2<q<1/2$, $f(z)$ is a holomorphic function for $Re z\geq
q$, $f(z)$ satisfies the condition
      \beq \label{1}
      \left| F(x+iy) \right|<\epsilon(x)e^{a|y|}, \quad 0<a<2\pi,
      \eeq
and $\epsilon(x)$ is an arbitrary function with asymptotic
behavior
      \beq
      \epsilon(x)\rightarrow 0 \quad \mbox{for} \quad x
      \rightarrow +\infty.
      \eeq
Using this formula we can calculate the sum (\ref{sun}):
      \bear \label{p1}
      S(u, n)&=&\lim _{q \rightarrow +0}\left\{
       \int\nolimits_{q}^{\infty}\left[\frac{x^{2n+1}}
       {(u^2+x^2)^{(2n-1)/2}}-x^2+(2n-1)\frac{u^2}{2}\right]dx
       \right. \nn && \left.+\int \nolimits_{q-i\infty}^{q}
       \left[ \frac{z^{2n+1}}{(u^2+z^2)^{(2n-1)/2}}
      -z^2+(2n-1)\frac{u^2}{2}\right]\frac{dz}
      {(1+e^{i2\pi z})} \right. \nn && \left.
      -\int \nolimits_{q}^{q+i\infty}\left[\frac{z^{2n+1}}
      {(u^2+z^2)^{(2n-1)/2}}-z^2+(2n-1)\frac{u^2}{2}\right]
      \frac{dz}{(1+e^{-i2\pi z})}  \right\} \nn &&
      =\frac{(2n-1)}{3}\frac{(2n)!! \ u^3}{(2n-1)!!}
      +2(-1)^m \lim_{\delta \rightarrow +0} \left[
      \int \nolimits_{0}^{u+\delta}\frac{x ^{2n+1}dx}
      {(u^2-x^2)^{(2n-1)/2}(1+e^{2\pi x})}\right. \nn && \left.
      -\left( { \mbox{\small {terms of this integral that}}
      \atop\mbox{\small{diverge in the limit}} \ \delta \rightarrow +0}
      \right)\right]\nn &&
      =\frac{(2n-1)\ u^3}{(2n-1)!!}\left\{\frac{(2n)!!}{3}
      -2\left(\frac{d}{u d u}\right)^n
      \left[u^{2n} \int \nolimits_{0}^{1}\frac{y \sqrt{1-y^2} dy}
      {1+e^{2\pi u y}} \right]\right\} \hspace{2cm} (n \geq 0).
      \ear
The sum in expression (\ref{37}) can be calculated by the
differentiation of $S(u,n)$:
      \bear \label{sunn}
      &&\sum \limits_{l=0}^{\infty}  \left\{\frac{\left(l+\frac12
      \right)^{2n+1}}{[u^2+\left(l+1/2\right)^2]^{(2n+1)/2}}
      -1 \right\}=\frac{-1}{(2n-1)}\left(\frac{d}{u du}\right) S(u, n)
      \nn && =-\frac{(2n)!! \ u}{(2n-1)!!} +\frac{2}{(2n-1)!!}
      \left(\frac{d}{u d u}\right)\left\{u^3
      \left(\frac{d}{u d u}\right)^n
      \left[u^{2n} \int \nolimits_{0}^{1}\frac{y \sqrt{1-y^2} dy}
      {1+e^{2\pi u y}} \right]
       \right\} \hspace{2cm} (n \geq 0).
      \ear
The other sums in the expression for $S_n^k(\varepsilon, u_0)$
[Eq. (\ref{intsums})] can also be calculated by the
differentiation of $S(u,n)$:
      \bear \label{sukn}
      &&\sum \limits_{l=0}^{\infty} \frac{\left(l+\frac12
      \right)^{2n+1}}{[u^2+\left(l+1/2\right)^2]^{(2n+2k+3)/2}}
      =\frac{(-1)^k (2n-1)!!}{(2n-1)(2n+2k+1)!!}
      \left(\frac{d}{u du}\right)^{k+2} S(u, n)\nn &&
      =-\frac{(2n)!!}{(2n+2k+1)!!}\frac{(2k-1)!!}{u^{2k+1}}
      \nn && -\frac{2 (-1)^k}{(2n+2k+1)!!}
      \left(\frac{d}{u d u}\right)^{k+2}\left\{u^3
      \left(\frac{d}{u d u}\right)^n
      \left[u^{2n} \int \nolimits_{0}^{1}\frac{y \sqrt{1-y^2} dy}
      {1+e^{2\pi u y}} \right]
       \right\} \hspace{2.3cm} \left(n \geq 0, \atop k \geq 0 \right).
      \ear
The integral over $y$ in all these expressions can be calculated
approximately. In this integral the integrand decreases
exponentially for $y > 1/(2 \pi u)$. The parameter $u$ in all
expressions (\ref{intsums}) satisfies the condition $u > u_0 \gg
1$. Therefore the main contribution to this integral gives a small
value of $y$ ($y \ll 1$) and the square root in the integrand can
be expanded. Then the up limit can be gone to infinity:
      \bear
      \int \nolimits_{0}^{1}\frac{y \sqrt{1-y^2} dy}
      {1+e^{2\pi u y}}&=&\int \nolimits_{0}^{\infty}\frac{y  dy}
      {1+e^{2\pi u y}}\left[1-\frac{y^2}{2}-\frac{1!!}{4!!}y^4
      -\frac{3!!}{6!!}y^6+O(y^{8})\right]
      \nn && =\frac{1}{48u^2}-\frac{7}{3840u^4}-\frac{31}{129024u^6}
      +O\left(\frac{1}{u^{10}} \right).
      \ear
Substituting this expression into Eqs. (\ref{p1})-(\ref{sukn}) and
integrating in Eq. (\ref{intsums}) we obtain the resulting
expressions (\ref{start})-(\ref{finish}).

\section*{Appendix B}

\setcounter{equation}{0}
\renewcommand{\theequation}{B\arabic{equation}}

      \bear
      \left(T^{t}_{t} \right)_{\!(4)}&=& \frac{1}{46080 \pi^2 r^8
f^4} \left\{
      360 r^8 f^4 m^4
      +\left[360 r^8 f^2 f'^2
      -480 r^6 f^3 f' {(r^2)}'
      -480 r^8 f^3 f''
\right]m^2              \right.\nn &&
      +32 r^4 f^4
      -32 {(r^2)}'''' r^6 f^4
      -12 f''^2 r^8 f^2
      +56 {(r^2)}''^2 r^4 f^4
      +16 f'''' r^8 f^3
      +7 f'^4 r^8                    \nn &&
      -32 f'' {(r^2)}'' r^6 f^3
      -16 f' {(r^2)}'^3 r^2 f^3
      +40 f'^2 {(r^2)}'' r^6 f^2
      -16 f' f''' r^8 f^2
      +16 f'' {(r^2)}'^2 r^4 f^3     \nn &&
      +4 f'^2 f'' r^8 f
      -12 f'^3 {(r^2)}' r^6 f
      -14 f'^2 {(r^2)}'^2 r^4 f^2
      -48 f' {(r^2)}''' r^6 f^3
      +48 {(r^2)}' {(r^2)}''' r^4 f^4  \nn &&
      -112 {(r^2)}'^2 {(r^2)}'' r^2 f^4
      +32 f' {(r^2)}'' {(r^2)}' r^4 f^3
      +40 {(r^2)}'^4 f^4
      +32 {(r^2)}' f''' r^6 f^3      \nn &&
      +\left[5760 r^4 f^4
      -11520 {(r^2)}'' r^4 f^4
      +2880 {(r^2)}'^2 r^2 f^4
      +8640 f'^2 r^6 f^2
      -11520 f'' r^6 f^3      \right.\nn &&
      +5760 {(r^2)}''^2 r^4 f^4
      +4320 f''^2 r^8 f^2
      +7560 f'^4 r^8
      +360 {(r^2)}'^4 f^4
      +11520 f'' {(r^2)}'' r^6 f^3 \nn &&
      +2880 f' {(r^2)}'^3 r^2 f^3
      -2880 f'^2 {(r^2)}'' r^6 f^2
      +5760 f' f''' r^8 f^2
      -2880 f'' {(r^2)}'^2 r^4 f^3 \nn &&
      -17280 f'^2 f'' r^8 f
      -11520 f'^3 {(r^2)}' r^6 f
      +720 f'^2 {(r^2)}'^2 r^4 f^2
      +11520 f' {(r^2)}''' r^6 f^3 \nn && \left.
      -2880 {(r^2)}'^2 {(r^2)}'' r^2 f^4
      +14400 f' f'' {(r^2)}' r^6 f^2
      -5760 f' {(r^2)}'' {(r^2)}' r^4 f^3
\right] \left(\xi-1/6 \right)^2 \nn &&
      +\left[\left(2880 r^6 f^4
      -5760 r^8 f^3 f''
      -5760 r^6 f^3 f' {(r^2)}'
      +4320 r^8 f^2 f'^2
      +720 r^4 f^4 {(r^2)}'^2 \right. \right.\nn && \left.
      -2880 r^6 f^4 {(r^2)}''\right) m^2
      -960 f'' r^6 f^3
      +960 {(r^2)}'' r^4 f^4
      +720 f'^2 r^6 f^2
      -960 {(r^2)}'^2 r^2 f^4 \nn &&
      -960 {(r^2)}'^4 f^4
      +1920 {(r^2)}' f''' r^6 f^3
      +960 {(r^2)}'''' r^6 f^4
      -2160 f''^2 r^8 f^2
      -1440 {(r^2)}''^2 r^4 f^4  \nn &&
      +960 f'''' r^8 f^3
      +2880 f'' {(r^2)}'' r^6 f^3
      +1200 f' {(r^2)}'^3 r^2 f^3
      -2160 f'^2 {(r^2)}'' r^6 f^2  \nn &&
      -2880 f' f''' r^8 f^2
      -1200 f'' {(r^2)}'^2 r^4 f^3
      +7920 f'^2 f'' r^8 f
      +2640 f'^3 {(r^2)}' r^6 f               \nn &&
      +900 f'^2 {(r^2)}'^2 r^4 f^2
      +1920 f' {(r^2)}''' r^6 f^3
      +2880 {(r^2)}'^2 {(r^2)}'' r^2 f^4
      -3780 f'^4 r^8                         \nn && \left.
      -1440 {(r^2)}' {(r^2)}''' r^4 f^4
      -2400 f' {(r^2)}'' {(r^2)}' r^4 f^3
      -4320 f' f'' {(r^2)}' r^6 f^2
\right]\left(\xi-1/6 \right)                 \nn &&
      + \left. \left[ -720m^4 r^8 f^4
      -16 r^4 f^4
      -20 {(r^2)}'^4 f^4
      +56 {(r^2)}'^2 {(r^2)}'' r^2 f^4
      +8 f' {(r^2)}'' {(r^2)}' r^4 f^3 \right. \right.\nn &&
      -28 {(r^2)}''^2 r^4 f^4
      -16 f'''' r^8 f^3
      +52 f' f'' {(r^2)}' r^6 f^2
      +16 {(r^2)}'''' r^6 f^4
      +36 f''^2 r^8 f^2             \nn &&
      +49 f'^4 r^8
      -4 f' {(r^2)}'^3 r^2 f^3
      -4 f'^2 {(r^2)}'' r^6 f^2
      +48 f' f''' r^8 f^2
      +4 f'' {(r^2)}'^2 r^4 f^3             \nn &&
      -116 f'^2 f'' r^8 f
      -22 f'^3 {(r^2)}' r^6 f
      -3 f'^2 {(r^2)}'^2 r^4 f^2
      +8 f' {(r^2)}''' r^6 f^3
      -32 {(r^2)}' f''' r^6 f^3              \nn &&
      -24 {(r^2)}' {(r^2)}''' r^4 f^4
      -8 f'' {(r^2)}'' r^6 f^3
      +\left(-2880 r^4 f^4
      +4320 {(r^2)}'^2 r^2 f^4
      +5580 {(r^2)}'^4 f^4         \right. \nn &&
      +5760 f' {(r^2)}'' {(r^2)}' r^4 f^3
      -15840 {(r^2)}'^2 {(r^2)}'' r^2 f^4
      +6480 f''^2 r^8 f^2
      +5760 {(r^2)}''^2 r^4 f^4            \nn &&
      -2880 f'''' r^8 f^3
      -5760 {(r^2)}'''' r^6 f^4
      -2880 f' {(r^2)}'^3 r^2 f^3
      +5760 f'^2 {(r^2)}'' r^6 f^2             \nn &&
      +15840 f' f'' {(r^2)}' r^6 f^2
      +8640 f' f''' r^8 f^2
      +2880 f'' {(r^2)}'^2 r^4 f^3
      -20880 f'^2 f'' r^8 f             \nn &&
      -9360 f'^3 {(r^2)}' r^6 f
      -2160 f'^2 {(r^2)}'^2 r^4 f^2
      -2880 f' {(r^2)}''' r^6 f^3
      -5760 {(r^2)}' f''' r^6 f^3             \nn && \left.
      +8640 {(r^2)}' {(r^2)}''' r^4 f^4
      -5760 f'' {(r^2)}'' r^6 f^3
      +8820 f'^4 r^8\right) \left(\xi-1/6\right)^2    \nn &&  \left. \left.
      +\left(2880 r^6 f^4 {(r^2)}''
      -720 r^4 f^4 {(r^2)}'^2
      -2880 r^6 f^4\right) \left(\xi-1/6\right) m^2
\right] \ln\left|\frac{4 u_0^2}{\mds^2 r^2 } \right|\right\},
      \ear

      \bear
      \left(T^{\rho}_{\rho} \right)_{\!(4)}&=&
      \frac{1}{46080 \pi^2 r^8f^4} \left\{
     -1080 m^4 r^8 f^4
     -120 {f'}^2 m^2 r^8 f^2
     +{f'}^4 r^8
     +32 {f'} {(r^2)}'' {{(r^2)}'} r^4 f^3 \right. \nn && \left.
     -4 {f'}^2 f'' r^8 f
     -24 f'' {{(r^2)}'}^2 r^4 f^3
     -16 {f'} {(r^2)}''' r^6 f^3
     -16 {{(r^2)}'} f''' r^6 f^3
     -8 {f'} {{(r^2)}'}^3 r^2 f^3    \right. \nn && \left.
     -24 {f'}^2 {(r^2)}'' r^6 f^2
     +12 {f'}^2 {{(r^2)}'}^2 r^4 f^2
     +8 {f'} f''' r^8 f^2
     -4 f''^2 r^8 f^2
     +16 f'' {(r^2)}'' r^6 f^3 \right. \nn && \left.
     -8 {f'}^3 {{(r^2)}'} r^6 f
     +32 {f'} f'' {{(r^2)}'} r^6 f^2
     +\left[5760 {f'} {(r^2)}'' {{(r^2)}'} r^4 f^3
     -1440 {f'} {{(r^2)}'}^3 r^2 f^3 \right. \right. \nn && \left. \left.
     -5760 {f'} {{(r^2)}'} r^4 f^3
     -2880 {f'}^2 r^6 f^2
     +2880 {f'} f'' {{(r^2)}'} r^6 f^2
     +2160 {f'}^2 {{(r^2)}'}^2 r^4 f^2 \right. \right. \nn && \left. \left.
     -720 {f'}^4 r^8
     +1440 {f'}^2 f'' r^8 f
     +2880 {f'}^2 {(r^2)}'' r^6 f^2\right] \left(\xi-1/6 \right)^2 \right. \nn && \left.
     +\left[-240 {f'}^2 r^6 f^2
     +\left(720 {{(r^2)}'}^2 r^4 f^4
     -1440 {f'} {{(r^2)}'} r^6 f^3
     -1440 {f'}^2 r^8 f^2 \right. \right. \right. \nn && \left. \left.
     \left.
     -2880 r^6 f^4  \right) m^2
     -1260 {f'}^2 {{(r^2)}'}^2 r^4 f^2
     +540 {f'}^4 r^8
     +480 {{(r^2)}'} f''' r^6 f^3
     -720 {f'}^2 f'' r^8 f \right. \right. \nn && \left. \left.
     +480 {f'} f''' r^8 f^2
     -480 f'' {(r^2)}'' r^6 f^3
     -720 {f'} {(r^2)}'' {{(r^2)}'} r^4 f^3
     -1200 {f'} f'' {{(r^2)}'} r^6 f^2 \right. \right. \nn && \left. \left.
     +1200 {f'}^2 {(r^2)}'' r^6 f^2
     +480 {f'} {(r^2)}''' r^6 f^3
     +720 {f'}^3 {{(r^2)}'} r^6 f
     +720 f'' {{(r^2)}'}^2 r^4 f^3 \right. \right. \nn && \left. \left.
     +480 {f'} {{(r^2)}'} r^4 f^3
     -240 f''^2 r^8 f^2\right] \left(\xi-1/6 \right)
+\left[-720 m^4 r^8 f^4
     +12 f'' {{(r^2)}'}^2 r^4 f^3 \right. \right. \nn && \left. \left.
     -16 r^4 f^4
     -4 {{(r^2)}'}^4 f^4
     +8 {f'}^2 {(r^2)}'' r^6 f^2
     +12 {f'}^2 f'' r^8 f
     +8 {{(r^2)}'} f''' r^6 f^3
     -8 {f'} f''' r^8 f^2 \right. \right. \nn && \left. \left.
     +8 {f'} {(r^2)}''' r^6 f^3
     -24 {f'} f'' {{(r^2)}'} r^6 f^2
     -3 {f'}^2 {{(r^2)}'}^2 r^4 f^2
     +4 {f'} {{(r^2)}'}^3 r^2 f^3 \right. \right. \nn && \left. \left.
     -16 {f'} {(r^2)}'' {{(r^2)}'} r^4 f^3
     +8 {{(r^2)}'}^2 {(r^2)}'' r^2 f^4
     -8 {{(r^2)}'} {(r^2)}''' r^4 f^4
     +10 {f'}^3 {{(r^2)}'} r^6 f \right. \right. \nn && \left. \left.
     -7 {f'}^4 r^8
     -8 f'' {(r^2)}'' r^6 f^3
     +4 f''^2 r^8 f^2
     +4 {(r^2)}''^2 r^4 f^4
     +\left(
     -2880 r^4 f^4 \right. \right. \right. \nn && \left. \left. \left.
     -4320 {{(r^2)}'}^2 r^2 f^4
     -2880 {f'}^2 {(r^2)}'' r^6 f^2
     -2340 {{(r^2)}'}^4 f^4
     -1440 {f'} f''' r^8 f^2  \right. \right. \right. \nn && \left. \left. \left.
     +2880 f'' {(r^2)}'' r^6 f^3
     +4320 {f'} f'' {{(r^2)}'} r^6 f^2
     -1440 {f'}^3 {{(r^2)}'} r^6 f
     +4320 {f'}^2 {{(r^2)}'}^2 r^4 f^2  \right. \right. \right. \nn && \left. \left. \left.
     +720 {f'} {{(r^2)}'}^3 r^2 f^3
     -2880 {f'} {(r^2)}''' r^6 f^3
     +5760 {{(r^2)}'}^2 {(r^2)}'' r^2 f^4
     -5760 {{(r^2)}'} {(r^2)}''' r^4 f^4 \right. \right. \right. \nn && \left. \left. \left.
     +1440 {f'} {(r^2)}'' {{(r^2)}'} r^4 f^3
     -1260 {f'}^4 r^8
     -2880 {{(r^2)}'} f''' r^6 f^3
     +2160 {f'}^2 f'' r^8 f \right. \right. \right. \nn && \left. \left. \left.
     -4320 f'' {{(r^2)}'}^2 r^4 f^3
     +720 f''^2 r^8 f^2
     +2880 {(r^2)}''^2 r^4 f^4  \right) \left(\xi-1/6 \right)^2
     +\left(-2880 r^6 f^4  \right. \right. \right. \nn && \left. \left. \left.
     +720 {{(r^2)}'}^2 r^4 f^4
     +1440 {f'} {{(r^2)}'} r^6 f^3 \right) \left(\xi-1/6 \right) m^2
\right] \ln\left|\frac{4 u_0^2}{\mds^2 r^2 }\right|
     \right\},
     \ear

      \bear
      \left(T^{\theta}_{\theta} \right)_{\!(4)}&=&
      \left(T^{\phi}_{\phi} \right)_{\!(4)}=
\frac{1}{46080 \pi^2 r^8 f^4} \left\{
      -1080 r^8 f^4 m^4
      +120 r^8 f^2 {f'}^2 m^2
      -8 r^4 f^3 {f'}{(r^2)}'' {(r^2)'}\right. \nn && \left.
      +16 r^6 f^2 {f'} f'' {(r^2)'}
      +17 r^8 {f'}^4
      -16 r^8 f^3 f''''
      -8 r^6 f {f'}^3 {(r^2)'}
      +16 r^6 f^3 {f'} {(r^2)}''' \right. \nn && \left.
      +16 r^4 f^3 f'' {(r^2)'}^2
      -52 r^8 f {f'}^2 f''
      +24 r^8 f^2 {f'} f'''
      -4 r^4 f^2 {f'}^2 {(r^2)'}^2\right. \nn && \left.
      +8 r^6 f^2 {f'}^2 {(r^2)}''
      +28 r^8 f^2 f''^2
      -24 r^6 f^3 {(r^2)'} f'''
      -16 r^6 f^3 f'' {(r^2)}'' \right. \nn && \left.
+\left[-5760 r^6 f^3 f''
      -1440 r^4 f^3 f'' {(r^2)'}^2
      +2880 r^6 f^2 {f'}^2
      -14400 r^4 f^3 {f'} {(r^2)}''{(r^2)'}\right. \right. \nn && \left.
      \left.
      +11520 r^6 f^3 {f'} {(r^2)}'''
      +2880 r^8 f^2 f''^2
      +5760 r^6 f^3f'' {(r^2)}''
      -7920 r^6 f {f'}^3 {(r^2)'}\right. \right. \nn && \left.
      \left.
      -3600 r^4 f^2 {f'}^2 {(r^2)'}^2
      -14400 r^8 f {f'}^2 f''
      +8640 r^4 f^3 {f'} {(r^2)'}
      +2880 r^6 f^2 {f'}^2 {(r^2)}''\right. \right. \nn && \left.
      \left.
      +5040 r^2 f^3{f'} {(r^2)'}^3
      +6480 r^8 {f'}^4
      +5760 r^8 f^2 {f'} f'''
      +10080 r^6 f^2 {f'} f'' {(r^2)'}\right] \left(\xi-1/6 \right)^2
      \right. \nn && \left.
+\left[-240 r^6 f^2 {f'}^2
      +480 r^6 f^3 f''
+\left(-720 r^4 f^4 {(r^2)'}^2
      -1440 r^8 f^3 f''
      +720 r^8 f^2 {f'}^2 \right. \right. \right. \nn &&
      \left. \left. \left.
      +1440 r^6 f^4 {(r^2)}''
      -720 r^6 f^3 {f'} {(r^2)'}\right) m^2
      -1440 r^8 f^2 f''^2
      -720 r^4 f^3 {f'} {(r^2)}'' {(r^2)'}\right. \right. \nn && \left.
      \left.
      -1920 r^8 f^2 {f'} f'''
      +5400 r^8 f {f'}^2 f''
      -2580 r^8 {f'}^4
      +720 r^6 f^3 f'' {(r^2)}''
      +480 r^8 f^3 f''''
      \right. \right. \nn && \left.\left.
      -2280 r^6 f^2 {f'} f'' {(r^2)'}
      +720 r^6 f^3 {(r^2)'} f'''
      -720 r^4 f^3 f'' {(r^2)'}^2
      +1440 r^6 f {f'}^3 {(r^2)'}\right. \right. \nn && \left.
      \left.
      +480 r^2 f^3 {f'} {(r^2)'}^3
      +660 r^4 f^2 {f'}^2 {(r^2)'}^2
      -960 r^6 f^2 {f'}^2 {(r^2)}''\right] \left(\xi-1/6 \right)
       \right. \nn && \left.
+\left[-720 r^8 f^4 m^4
      +16 r^4 f^4
      -20 r^8 f^2 {f'} f'''
      +12 f^4 {(r^2)'}^4
      -14 r^6 f^2 {f'} f'' {(r^2)'} \right. \right. \nn && \left. \left.
      +12 r^6 f^3 {(r^2)'} f'''
      +8 r^6 f^3 f'' {(r^2)}''
      +6 r^6 f {f'}^3 {(r^2)'}
      -2 r^6 f^2 {f'}^2 {(r^2)}''
      -8 r^6 f^3 {f'} {(r^2)}''' \right. \right. \nn && \left. \left.
      +12 r^4 f^4 {(r^2)}''^2
      -20 r^8 f^2 f''^2
      +4 r^4 f^3 {f'} {(r^2)}'' {(r^2)'}
      -8 r^4 f^3 f'' {(r^2)'}^2
      +3 r^4 f^2 {f'}^2 {(r^2)'}^2 \right. \right. \nn && \left. \left.
      +16 r^4 f^4 {(r^2)'} {(r^2)}'''
      +52 r^8 f {f'}^2 f''
      +8 r^8 f^3 f''''
      -8 r^6 f^4 {(r^2)}''''
      -32 r^2 f^4 {(r^2)'}^2 {(r^2)}'' \right. \right. \nn && \left. \left.
  +\left(2880 r^4 f^4
      -21 r^8 {f'}^4
      -8640 r^4 f^4 {(r^2)}''
      -4320 r^4 f^3 {f'} {(r^2)'}
      +8640 r^2 f^4 {(r^2)'}^2\right. \right. \right. \nn && \left. \left.
      \left.
      +9360 r^6 f^2 {f'} f'' {(r^2)'}
      +7200 r^8 f^2 {f'} f'''
      -3240 r^4 f^2 {f'}^2 {(r^2)'}^2
      +5040 r^6 f^2 {f'}^2 {(r^2)}'' \right. \right. \right. \nn && \left. \left.
      \left.
      +7020 f^4 {(r^2)'}^4
      -5400 r^2 f^3 {f'} {(r^2)'}^3
      -20880 r^2 f^4 {(r^2)'}^2 {(r^2)}''
      +11520 r^4 f^4 {(r^2)'} {(r^2)}''' \right. \right. \right. \nn && \left. \left.
      \left.
      -16560 r^8 f {f'}^2 f''
      -5400 r^6 f {f'}^3 {(r^2)'}
      +8640 r^4 f^4 {(r^2)}''^2
      +11520 r^4 f^3 {f'} {(r^2)}'' {(r^2)'} \right. \right. \right. \nn && \left. \left.
      \left.
      -5760 r^6 f^3 {f'} {(r^2)}'''
      -7200 r^6 f^3 f'' {(r^2)}''
      -4320 r^6 f^3 {(r^2)'} f'''
      -2880 r^8 f^3 f'''' \right. \right. \right. \nn && \left. \left.
      \left.
      +5040 r^8 f^2 f''^2
      +5040 r^4 f^3 f'' {(r^2)'}^2
      -5760 r^6 f^4 {(r^2)}''''
      +7020 r^8 {f'}^4 \right) \left(\xi-1/6 \right)^2 \right. \right. \nn && \left. \left.
  +\left(-720 r^4 f^4 {(r^2)'}^2
      +720 r^6 f^3 {f'} {(r^2)'}
      +1440 r^6 f^4{(r^2)}''
      +1440 r^8 f^3 f''  \right. \right. \right. \nn && \left. \left.
      \left.
      -720 r^8 f^2 {f'}^2\right) \left(\xi-1/6 \right) m^2
      \right] \ln\left|\frac{4 u_0^2}{\mds^2 r^2 }\right|
\right\}.
      \ear


\end{document}